\documentclass{ws-ijmpd}
\begin{document}

\markboth{Xing Hu Li, Zhi Fu Gao, Xiang Dong Li, Yan Xu, Pei Wang, and Na Wang
}{Numerically Fitting The Electron Fermi Energy and The Electron Fraction in A Neutron Star}

\catchline{}{}{}{}{}

\title{Numerically Fitting The Electron Fermi Energy and The Electron Fraction in A Neutron Star }

\author{Xing Hu, Li}
\address{1. Xinjiang Astronomical Observatory, CAS, 150, Science
1-Street, Urumqi, Xinjiang, China. \\
2. Key Laboratory of Radio Astronomy, Chinese Academy of Sciences, 2 West Beijing Road, Nanjing, Jiangsu, 210008, China \\
3. University of Chinese Academy of Sciences, 19A Yuquan Road, Beijing, 100049, China }
\author{Zhi Fu, Gao\footnote{zhifugao@xao.ac.cn}}
\address{1. Xinjiang Astronomical Observatory, CAS, 150, Science
1-Street, Urumqi, Xinjiang, China \\
2. Key Laboratory of Radio Astronomy, Chinese Academy of Sciences, 2 West Beijing Road, Nanjing, Jiangsu, 210008, China}
\author{Xiang Dong, Li}
\address{Shchool of Astronomy and Space Science, Nanjing University, Nanjing, Jiangshu, China}
\author{ Yan, Xu}
\address{Changchun Astronomical Observatory, National Observatories, CAS, Changchun, 130117, China}
\author{ Pei, Wang}
\address{National Astronomical Observatories, Chiese Academy of Sciences, Beijing 100012, China}
\author{Na, Wang}
\address{Xinjiang Astronomical Observatory, CAS, 150, Science
1-Street, Urumqi, Xinjiang, China}
\author{Jianping, Yuan}
\address{Xinjiang Astronomical Observatory, CAS, 150, Science
1-Street, Urumqi, Xinjiang, China}
\maketitle

\begin{history}
\received{Day Month Year} \revised{Day Month Year}
%\accepted{Day Month Year}
%\comby{(xxxxxxxxxx)}
\end{history}

\begin{abstract}
Based on the basic definition of Fermi energy of degenerate and
 relativistic electrons, we obtain a special solution to electron
Fermi energy, $E_{\rm F}(e)$, and express $E_{\rm F}(e)$ as a function
of electron fraction, $Y_{e}$, and matter density, $\rho$. Several useful
analytical formulae for $Y_{e}$ and $\rho$ within classical models and the work of Dutra et al. 2014
(Type-2) in relativistic mean field theory are obtained using numerically
fitting. When describing the mean-field Lagrangian,
density, we adopt the TMA parameter set, which is remarkably consistent with
with the updated astrophysical observations of neutron stars.  Due to the importance of the
density dependence of the symmetry energy, $S$, in nuclear astrophysics, a brief discussion on
the symmetry parameters $S_v$ and $L$ (the slope of $S$) is presented.
Combining these fit formulae with boundary conditions for different density
regions, we can evaluate the value of $E_{\rm F}(e)$ in any given matter
density, and obtain a schematic diagram of $E_{\rm F}(e)$ as a continuous
function of $\rho$.  Compared with previous study on the electron Fermi
energy in other models, our methods of calculating $E_{\rm F}(e)$ are
more simple and convenient, and can be universally suitable for the
 relativistic electron regions in the circumstances of common neutron stars.
We have deduced a general expression of $E_{\rm F}(e)$ and $n_{e}$, which could be used to
indirectly test whether one EoS of a NS is correct in our future
studies on neutron star matter properties. Since URCA reactions are expected in the
center of a massive star due to high-value electron Fermi energy and electron fraction,
 this study could be useful in the future studies on the NS thermal evolution.

\end{abstract}
%\ccode{PACS: 97.60.Jd; 21.65.-f;71.18.+y}
%\pacs{98.80.Es, 98.80.-k, 95.36.+x}
\section{Introduction}\label{I}
Pulsars are among the most mysterious objects in the universe
that provide natural laboratory for investigating the nature of matter
under extreme conditions, and are universally recognized as normal neutron
stars (NSs), but sometimes have been argued to be quark stars\cite{Xu02,Du09,Lai13} . The equation of
state (EoS) of matter under exotic conditions is an important tool for the
understanding of the nuclear force and for astrophysical applications. The
Fermi energy of relativistic electrons $E_{\rm F}(e)$ is one of most
important and indispensable physical parameters in the EoS, and varies with matter density $\rho$.
 The electron Fermi energy influences directly weak-interaction
processes including modified URCA reactions\cite{Yakovlev01},
electron capture\cite{Gao11a,Gao11b,Gao12a,Wang12,Liu13} and so on,
in the circumstance of a NS. These
influences will change intrinsic EoS, interior structure and heat evolution, and even affect the whole properties of the star.

Another important physical parameter affecting internal properties of
a NS is the electron fraction, which is defined as $Y_{e}=n_{e}/n_{B}$ ($n_{e}$ and $n_{B}$ are the electron
number density, baryon number density, respectively, and varies with the high-density star matter. The value
of average electron fraction is about 0.05 for a common
NS\cite{Yakovlev01,Shapiro83}. However, to exactly
calculate the values of $Y_{e}$ for the neutron star matter has long
been a very challenging task for both the nuclear physics and the
astrophysics community\cite{Lattimer04} due to some
uncertainties and artificial assumptions in studying structures and
properties of NSs. Currently, our knowledge of the electron fraction
mainly comes from the model-dependence EoSs of NSs. For the $\beta$-
equilibrated NS matter we have free neutron decay
$n\rightarrow p+ e^{-}+ \overline{\nu_{e}}$, which are governed by
weak interaction and the electron capture process $p+e^{-}\rightarrow n+ \nu_{e}$.
Both types of reactions alter $Y_e$, and thus affect the EoS. However,
in the currently popular and reliable EoSs of NSs, neutrinos generated in
these reactions are always assumed to leave the system. The absence of
neutrino surely has some influence on the EoS and mainly induces a significant
change on the value of $Y_e$.

As we know, for degenerate and relativistic electrons in $\beta-$equilibrium,
the distribution function $f(E_{e})$ obeys Fermi-Dirac statistics:
$f(E_{e})=1/(Exp((E_{e}-\mu_{e})/kT)+1)$, $k$ represents Boltzmann's
constant, and $\mu_{e}$ is the electron chemical potential. If $T\rightarrow 0$,
$\mu_{e}$ is also called the electron Fermi
energy, $E_{\rm F}(e)$, which presents the energy of highest occupied states for electrons.
The electron Fermi energy $E_{\rm F}(e)$ has the simple form
\begin{equation}
E_{\rm F}(e) = \left(p^{2}_{\rm F}(e)c^{2}+ m^{2}_{e}c^{4}\right)^{1/2}~~,~~%
\label{1}
\end{equation}
with $p_{\rm F}(e)$ being the electron Fermi momentum. The isoenergetic surface
of $E=E_{\rm F}(e)$ is called the electron Fermi surface\cite{Gao12b,Gao13,Gao11c}.

In the context of general relativity principle, the matter density is defined as:
$\rho= \varepsilon/c^{2}$, $\varepsilon$ is the total energy density, including the
rest-mass energies of particles. Using the basic thermodynamics, we obtain the
relation of the total matter pressure $P$ and matter density $\rho$ in a common NS,
\begin{eqnarray}
&&P(n_{B})= n_{B}^{2}\frac{d(\varepsilon/n_{B})}{d n_{B}}~, \nonumber\\
&&\rho(n_{B})= \varepsilon(n_{B})/c^{2}, ~~\Rightarrow P= P(\rho)~. % 2
\label{2}
\end{eqnarray}
From the above equation, it is obvious
that $P$ solely depends on $\rho$. Theoretically, we can obtain the value of $E_{\rm F}(e)$ by
solving EOS in a specific matter model. In this paper, we mainly focus on $E_{\rm F}(e)$ and $Y_e$ for degenerate and
relativistic electrons inside a common NS, where the magnetic effects
on the EoS are ignored.

This paper is arranged as follows. In Sec. 2, we briefly review the structure of a NS. In Sec. 3, we deduce a
special solution to $E_{\rm F}(e)$.  We apply the special solution obtained to numerically simulate
$E_{\rm F}(e)$ and $Y_e$ of a NS within classical models in Sec. 4, and within the work of Dutra
et al. 2014\cite{Dutra14} in Sec. 5. In Sec. 6 we present comparisons and discussions. The conclusion is presented in Sec.7.

\section{Reviewing A NS' Structure }\label{II}
The structure of a NS includes an atmosphere and four main internal
regions: the outer crust, inner crust, outer core, and inner core. The atmosphere
is a thin layer of plasma which determines
the spectrum of thermal electromagnetic radiation of
the star. The geometrical depth of the atmosphere varies
from some ten centimeters in a hot star down to some millimeters
in a cold one.

The knowledge of structures and properties of the crust plays an
important role in understanding many astrophysical
observations\cite{Horowitz04,Owen05,Steiner05,Burrows06}. The outer
crust consists of nuclei in a Coulomb lattice and electron gas, and has
a depth of a few hundred meters.  At $\rho\leq 10^{4}$~g~cm$^{-3}$, the electron gas may be
non-degenerate, and the ionization may be incomplete; at $\rho<10^{7}$~g~cm$^{-3}$,
the ground state may be $^{56}_{26}Fe$; at $\rho\sim10^{9}$~g~cm$^{-3}$,
the nuclei capture electrons and become neutron-rich, and at the neutron drip
density $\rho_{d}$, the neutrons start to drip from the nuclei and form a free neutron gas.
The neutron drip density $\rho_{d}$ is well determined\cite{BPS71,Ruster06} to be about
$4.3\times10^{11}$~g~cm$^{-3}$.

The inner crust of a NS spans the region from the neutron drip
point to the inner edge separating the solid crust from the
liquid core, and has several kilometers deep. The matter of the inner crust in nuclear
equilibrium consists of neutron-rich nuclei in a Coulomb lattice, electrons,
and free neutrons. As matter density increases, free neutrons supply an
increasingly large fraction of the total matter pressure $P$. At the crust-core
transition density, $\rho_{t}$, the baryon number density reaches a critical value, and
nuclei dissolve and merge together.

The outer core consists of neutrons, protons and electrons
(hereinafter ``$npe$ system''), spans a density range
of about $0.5\rho_{0}\sim 2.5\rho_{0}$, and has a depth of several
kilometers, $\rho_{0}=2.8\times10^{14}$~g~cm$^{-3}$ is the
standard nuclear density.  At the inner edge of the outer core
small fraction of muons ($\mu$) may appear\cite{Yakovlev01},
because the electron Fermi energy could exceed the muon rest-mass energy $m_{\mu}c^2=105.7$ MeV.
However, a definite boundary between the outer core and inner core has not yet been obtained due to the
uncertainty of crust-core transition density.

The inner core is about several kilometers
in radius, and has a central density as high as $\sim$
$10^{15}$~g~cm$^{-3}$. With still further increase of
density, the inner core becomes energetically more economic if some
nucleons transform to ¡®exotic¡¯ particles such as hyperons,
pion condensates, kaon condensates and quarks, etc., when
$\rho>\rho_{tr}$, here $\rho_{tr}$ is the transition density to these
¡®exotic¡¯ particles, which is $\sim 4\rho_{0}$\cite{Tsuruta09}.
The maximum of the inner core density could exceed
this transition density, so hyperons, pion condensates, kaon
condensates, quarks and nucleons with large $Y_e$ are expected
to exist in the inner core of a NS.

\section{Electron Fermi Energy And Its Special Solution}\label{III}
\subsection{General relation of $E_{\rm F}(e)$, $Y_e$ and $\rho$}
In the interior of a common NS with $B\sim 10^{10}-10^{12}$~G,
the isoenergetic surface of degenerate and relativistic electrons
is a spherical surface, and the microscopic state number of
electrons in a unit volume, $N_{pha}$, is calculated by
\begin{equation}
N_{pha}=n_{e}= \frac{g}{h^{3}}\int_{0}^{p_{\rm F}(e)}4\pi p^{2}dp=
\frac{8\pi}{3h^{3}}p_{\rm F}^{3}(e).
\label{3}
\end{equation}
where $h$ is Plank's constant. For the convenience of calculations, we introduce a dimensionless momentum of
electrons, $x_{e}=p_{\rm F}(e)/m_{e}c$. According to Pauli¡¯s exclusion principle, the
electron number density is equal to its microscopic state density,
\begin{equation}
n_{e}= N_{pha}= \frac{8\pi}{3\lambda_{e}^{3}}x_{e}^{3}~,
\label{4}
\end{equation}
where~$\lambda_{e}=h/m_{e}c= 2.4263\times 10^{-10}$~cm is the electron Compton wavelength.
The average mass of a baryon $m_{B}$~is defined as
\begin{equation}
m_{B}\equiv \frac{1}{n}\sum_{i}n_{i}m_{i}= \frac{\sum_{i}n_{i}m_{i}}{\sum_{i}n_{i}A_{i}},
\label{5}
\end{equation}
with~$A_{i}$~ the baryon number of species $i$. In the interior of a NS, the relation of
$m_{B}\equiv m_{u}\equiv1.6606\times10^{-24}$~g~always holds, $m_{u}$ is the mass of an atom.
Thus, the matter density can be expressed as:
\begin{equation}
\rho= n_{B}m_{B}= \frac{n_{e}m_{B}}{Y_{e}}~~,
\label{6}
\end{equation}
Combining Eq.(5) with Eq.(6), we get
\begin{equation}
x_{e}=(\frac{3\lambda_{e}^{3}}{8\pi m_{u}}Y_{e}\rho )^{\frac{1}{3}}~~.
\label{7}%
\end{equation}
Inserting~the values of $\lambda_{e}$ and $m_{u}$ into Eq.(7) gives
\begin{equation}
x_{e}=1.0088\times 10^{-2}(Y_{e}\rho)^{\frac{1}{3}}~~.
\label{8} %
\end{equation}
Since the average molecule weight of electrons~$\mu_{e}=\frac{m_{B}}
{m_{u}Y_{e}}=\frac{1}{Y_{e}}$, we get
\begin{equation}
\rho=\mu_{e}m_{u}n_{e}= 0.97395\times 10^{6}\frac{x_{e}^{3}}{Y_{e}},
\label{9} %
\end{equation}
Combining Eq.(1) with Eqs.(8) and (9), we get the electron Fermi energy
\begin{equation}
E_{\rm F}(e)=m_{e}c^{2}(1+ x_{e}^{2})^{1/2}=[1+1.018\times10^{-4}(\rho Y_{e})^{\frac{2}{3}}]
^{\frac{1}{2}}\times 0.511~{\rm MeV}.~
 \label{10} ~
\end{equation}
The expression above is a general formula for the electron Fermi energy,
which is approximately suitable for relativistic electron
matter regions ($\rho\geq 8.6\times 10^{6}$~g~cm$^{-3}$) in the whole interior
of a common NS. If $x_{e}=p_{e}/m_{e}c\gg 1 $, Eq.(10)
 can be approximately reduced as $E_{\rm F}(e)\approx 5.16\times10^{-3}(\rho Y_{e})^{\frac{1}{3}}$~{\rm MeV}.

\subsection{Special solution to $E_{\rm F}(e)$ }
 In this part, in order to obtain a special solution to
 $E_{\rm F}(e)$, we consider an ideal $n-p-e$ system
 under $\beta$-equilibrium in the outer core, where electrons are relativistic,
neutrons and protons are non-relativistic. According to Shapiro \& eukolsky (1983)\cite{Shapiro83}
(hereinafter ``ST-83''), when $\rho \gg 10^{13}$
~g~cm$^{-3}$, the neutron pressure dominates in the interior of a NS, and $\rho
\approx m_{n}n_{n}$, then the neutron number density $n_{n}=1.7\times 10^{38}
(\rho/\rho_{0})$~cm$^{-3}$, the charge neutrality requires the proton number
density $n_{p}=n_{e}=9.6\times 10^{35}(\rho/\rho_{0})^{2}$~cm$^{-3}$;
$\beta$-equilibrium implies momentum conservation of $p_{\rm F}(p)=p_{\rm F}
(e)= 60(\rho/\rho_{0})^{2/3}$~MeV/c), thus the electron Fermi energy
\begin{equation}
E_{\rm F}(e)= 60(\rho/\rho_{0})^{2/3}~~{\rm MeV},
\label{11}
\end{equation}
where $p_{\rm F}(p)$ is the proton Fermi momentum.  A simple proof of Eq.(11)
is presented as follows: From the relations of $n_{p}$, $n_{e}$ and $n_{n}$ above, we get
\begin{equation}
Y_{e}=Y_{p}\approx \frac{n_{e}}{n_{n}}\approx 0.005647\times(\frac{\rho}{\rho_{0}})~,
\label{12}
\end{equation}
where $Y_{p}$ is the proton fraction.
Combining Eq.(12) with Eq.(10), we get
\begin{eqnarray}
&&E_{\rm F}(e)=[1+1.018\times10^{-4}(\rho\times0.005647(\frac{\rho}{\rho_{0}}))^{\frac{2}{3}}]^{\frac{1}{2}}\times0.511~ \nonumber\\
&&=[1+1.018\times0^{-4}(2.8\times10^{14}\times0.005647(\frac{\rho}{\rho_{0}})^{2})^{\frac{2}{3}}]^{\frac{1}{2}}\times0.511~ \nonumber\\
&&=[1+1.018\times 10^{-4}(1.58116\times 10^{12})^{\frac{2}{3}}]^{\frac{1}{2}}\nonumber\\
&&\times(\frac{\rho}{\rho_{0}})^{\frac{2}{3}}\times0.511~
=60\times(\frac{\rho}{\rho_{0}})^{\frac{2}{3}}~({\rm MeV}).
\label{13} %
\end{eqnarray}
The above proof indicates that Eq.(10) is an accurate expression.
Inserting Eq.(12) into Eq.(11) yields
\begin{eqnarray}
E_{\rm F}(e)=&&60\times(\frac{\rho}{\rho_{0}})^{\frac{1}{3}}(\frac{\rho}{\rho_{0}})^{\frac{1}{3}}= 60\times(\frac{\rho}{\rho_{0}})^{\frac{1}{3}}(\frac{\frac{\rho_{0}Y_{e}}{0.005647}}{\rho_{0}})^{\frac{1}{3}}\nonumber\\
&& =60\times(\frac{\rho}{\rho_{0}})^{\frac{1}{3}}(\frac{Y_{e}}{0.005647})^{\frac{1}{3}}~~~~({\rm MeV}).
\label{14}
\end{eqnarray}
The equation above is a special solution to $E_{\rm F}(e)$, which is suitable relativistic electron
matter region of a common NS.

\subsection{Test the validity of the special solution }
To test the validity of
the special solution to $E_{\rm F}(e)$, we will calculate the values of
$E_{\rm F}(e)$ in the work of Baym, Bethe \&
Pethick (1971)\cite{BBP71} (hereinafter ``BBP model'') by using Eq.(14), and will compare our
results with those of BBP model. By introducing a compressible liquid drop model
of nuclei, BBP model is more successful than other models in describing matter in
the inner crust, where the system becomes a mixture of nuclei, free neutrons, and electrons.
According to BBP model\cite{BBP71},  the total energy density $\varepsilon$ and
the total matter pressure $P$ of the system are described by
\begin{eqnarray}
&&\varepsilon= \varepsilon_{e}(n_{e})+ n_{N}(W_{N} + W_{L})
+ n_{n}(1-~V_{N}n_{N})W_{n}~~,\nonumber\\
&& P= P_{n} + P_{e} + P_{L}~
\label{16}~
\end{eqnarray}
where $n_{n}$ is the number density of neutrons outside of nuclei
(hereinafter ``neutron gas''), and the new feature is the dependence on the
volume of a nucleus $V_{N}$, which decreases with the outside pressure
of the neutron gas; $P_{n}$, $P_{e}$ and $P_{L}$ are the neutron gas
pressure, electron pressure, and lattice pressure, respectively.
The baryon number density in this model is
\begin{equation}
n_{B}= An_{N} + (1- V_{N}n_{N})n_{n}~~~,
\label{16}
 \end{equation}
where $V_{N}n_{N}$ and $1- V_{N}n_{N}$ are the fraction of
volume occupied by nuclei, and the fraction occupied by neutron gas,
respectively. Then $E_{\rm F}(e)$ is determined by
\begin{equation}
E_{\rm F}(e)=\frac{\partial \varepsilon_{e}}{\partial n_{e}}= \frac{\partial}{\partial n_{e}}(n_{e}E_{e})
=- \frac{\partial}{\partial Z}(E_{N}+E_{L}).
~\label{17}~
 \end{equation}
The values of $E_{\rm F}(e)$ in BBP model are partly listed in Table 1. The sign `$\dag$' denotes that $\rho_{m}$ is
the maximum equilibrium density at which the nuclide is present.
The data of columns 1, 2, 3, 4 and 7 are cited from Table 2 of Canuto
(1974)\cite{Canuto74}.  The electron fraction $Y_{e}
=Y_{p}= \frac{Zn_{N}I}{n_{B}}\simeq \frac{Zn_{N}}{\rho/m_{u}}$, and
the values of $E_{\rm F}^{'}(e)$ in column 8 are obtained using the
specific solution of Eq.(14).
\begin{table*}[htb]
\tbl{Calculations of $Y_{e}$ and $E_{\rm F}(e)$ in BBP model.}
{\begin{tabular}{@{}cccccccc@{}} \toprule
$\rho_{m}^{\dag}$        & $A$ & $Z$ & $n_{N}$& $n_{B}$   &$Y_{e}$ & $E_{\rm F}(e)$& $E_{\rm F}^{'}(e)$ \\
g~cm$^{-3}$ &   &      &  $10^{-6}$fm$^{-3}$ & cm$^{-3}$   &           &MeV & MeV \\
\colrule
4.66$\times 10^{11}$&127 &40 &2.02 &2.806$\times 10^{35}$  &0.2879 &26.31 & 26.36\\
6.61$\times 10^{11}$&130 &40 &2.13 &3.981$\times 10^{35}$  &0.2140 &26.98 & 26.83\\
8.79$\times 10^{11}$&134 &41 &2.23 &5.293$\times 10^{35}$  &0.1727 &27.51 & 27.47\\
1.20$\times 10^{12}$&137 &42 &2.34 &7.226$\times 10^{35}$  &0.1360 &28.13 & 28.14 \\
1.47$\times 10^{12}$&140 &42 &2.43 &8.852$\times 10^{35}$  &0.1153 &28.58 & 28.51\\
2.00$\times 10^{12}$&144 &43 &2.58 &1.204$\times 10^{36}$  &0.0921 &29.33 & 29.30\\
2.67$\times 10^{12}$&149 &44 &2.74 &1.608$\times 10^{36}$  &0.0749 &30.15 & 30.12 \\
3.51$\times 10^{12}$&154 &45 &2.93 &2.114$\times 10^{36}$  &0.0624 &31.05 & 31.04\\
4.54$\times 10^{12}$&161 &46 &3.14 &2.734$\times 10^{36}$  &0.0528 &32.02 & 32.00\\
6.25$\times 10^{12}$&170 &48 &3.45 &3.764$\times 10^{36}$  &0.0439 &33.43 & 33.47\\
8.38$\times 10^{12}$&181 &49 &3.82 &5.046$\times 10^{36}$  &0.0371 &34.98 & 34.90\\
1.10$\times 10^{13}$&193 &51 &4.23 &6.624$\times 10^{36}$  &0.0326 &36.68 &  36.59\\
1.50$\times 10^{13}$&211 &54 &4.84 &9.033$\times 10^{36}$  &0.0289 &39.00 & 38.98 \\
1.99$\times 10^{13}$&232 &57 &5.54 &1.198$\times 10^{37}$  &0.0264 &41.58 & 41.56\\
2.58$\times 10^{13}$&257 &60 &6.36 &1.554$\times 10^{37}$  &0.0246 &44.37 & 44.26 \\
3.44$\times 10^{13}$&296 &65 &7.52 &2.071$\times 10^{37}$  &0.0236 &48.10 & 48.04  \\
4.68$\times 10^{13}$&354 &72 &9.12 &2.818$\times 10^{37}$  &0.0233 &52.95 & 53.01 \\
5.96$\times 10^{13}$&421 &78 &10.7 &3.589$\times 10^{37}$  &0.0235 &57.56 & 57.62 \\
8.01$\times 10^{13}$&548 &89 &13.1 &4.824$\times 10^{37}$  &0.0242 &64.32 &  64.21 \\
9.83$\times 10^{13}$&683 &100 &15.0 &5.920$\times 10^{37}$ &0.0253 &69.81 & 69.78 \\
1.30$\times 10^{14}$&990 &120 &17.8 &7.828$\times 10^{37}$ &0.0273 &78.58 &  78.56\\
1.72$\times 10^{14}$&1640 &157 &19.6 &1.035$\times10^{38}$&0.0297 &88.84 & 88.70 \\ \botrule
\end{tabular} \label{ta1}}
\end{table*}

In Table 1 the magnitude of $|\frac{E_{\rm F}^{'}(e)-E_{\rm F}(e)}
{E_{\rm F}(e)}|$ is $\sim 10^{-3}$, which
illustrates that our calculations are consistent with those of BBP\cite{BBP71}.
From the comparison above, our method of calculating
$E_{\rm F}(e)$ is more simple and convenient than that in BBP model.
Be noted that, as a representative model, BBP model is arbitrarily selected
to test the validity of the specific solution to $E_{\rm F}(e)$. In this part,
other matter models will no longer be enumerated, due to the restriction of space.

\section{Numerically Simulating $E_{\rm F}(e)$ and $Ye$ in A NS (I)}\label{IV}
In this Section, we will
numerically fit the relation of $Y_{e}$ and $\rho$ in the whole
interior of a NS, according to several reliable EoSs within simple but classical matter models.
Combining the analytic expressions of $Y_{e}$ and $\rho$ with the special
solution to $E_{\rm F}(e)$, we will obtain schematic diagrams of
$E_{\rm F}(e)$ vs. $\rho$ in different density regions for relativistic electrons.
\subsection{Numerically fitting in the outer crust}
 By introducing the lattice energy, Baym, Pethick \& Sutherland (1971)\cite{BPS71}
(hereafter ``BPS model'') improved on Salpeter's treatment\cite{Salpeter61}, and described the nuclear
composition and EoS for catalyzed matter in complete thermodynamic
equilibrium below $\rho_{d}$. BPS model is one of most successful models describing matter of the outer crust.
According to BPS model, the total energy density $\varepsilon$ and
the total matter pressure $P$ are described by
\begin{eqnarray}
&&\varepsilon=\varepsilon_{N}+\varepsilon_{e}+\varepsilon_{L},\nonumber\\
&&P= P_{e} + P_{L}= P_{e}+ \frac{1}{3}\varepsilon_{L},
  \label{18}
\end{eqnarray}
where $\varepsilon_{N}$ is the energy density of nucleus, $\varepsilon_{e}$
the energy density of free electrons, and $\varepsilon_{L}$ the
$bcc$ Coulomb lattice energy in a unit volume.  The value of
$E_{\rm F}(e)$ is obtained by solving the following differential equation,
\begin{equation}
E_{\rm F}(e)=\mu_{e}=\frac{\partial \varepsilon_{e}}{\partial n_{e}}
=\frac{\partial}{\partial n_{e}}( n_{e}E_{e}).
\label{19}
\end{equation}
For the specific equilibrium nuclei $(A, Z)$ in BPS model, the
values of quantities $Y_{e}$, $\rho_{m}$, and $E_{\rm F}(e)$
are tabulated in Table 2.
\begin{table*}[htb]
\tbl{Values of $Y_{e}$, $\rho_{m}$, and $E_{\rm F}(e)$
in BPS model below neutron drop.}
{\begin{tabular}{@{}cccccccc@{}} \toprule
Nuclei & $Y_{e}$& $\rho_{m}$&  $E_{\rm F}(e)$ & Nuclei & $Y_{e}$& $\rho_{m}$&  $E_{\rm F}(e)$\\
        &    & g~cm$^{-3}$  & MeV  &  &  & g~cm$^{-3}$  & MeV \\
\colrule
$^{56}_{26}Fe$ & 0.4643 &8.1$\times 10^{6}$  &0.95 & $^{78}_{28}Ni$ & 0.3590 &8.21$\times 10^{10}$ & 20.0 \\
$^{62}_{28}Ni$ & 0.4516 &2.7$\times 10^{8}$  &2.60  & $^{76}_{26}Fe$& 0.3421 &1.8$\times 10^{11}$ &20.20\\
$^{64}_{28}Ni$ & 0.4375&1.2$\times 10^{9}$ &4.20  & $^{124}_{42}Mo$& 0.3387 &1.9$\times 10^{11}$ &20.50 \\
$^{84}_{34}Se$ & 0.4048 &8.2$\times 10^{9}$ &7.70  & $^{122}_{40}Zr$& 0.3279 &2.7$\times 10^{11}$ &22.90 \\
$^{82}_{32}Ge$ & 0.3902 &2.2$\times 10^{10}$ &10.60 & $^{120}_{38}Sr$& 0.3167 &3.7$\times 10^{11}$ & 25.2\\
$^{80}_{30}Zn$ & 0.3750 &5.91$\times 10^{10}$ &13.60 & $^{118}_{36}Kr$& 0.3051 &4.3$\times 10^{11}$ &26.20\\
 \botrule
\end{tabular} \label{ta2}}
\end{table*}

\begin{figure*}[htb]
\begin{center}
\begin{tabular}{cc}
\scalebox{0.65}{\includegraphics{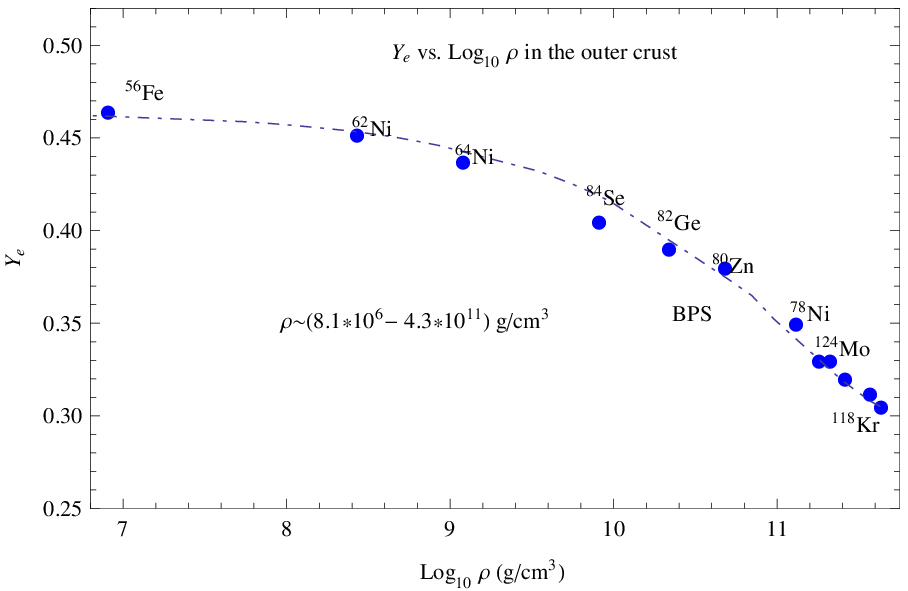}}&\scalebox{0.65}{\includegraphics{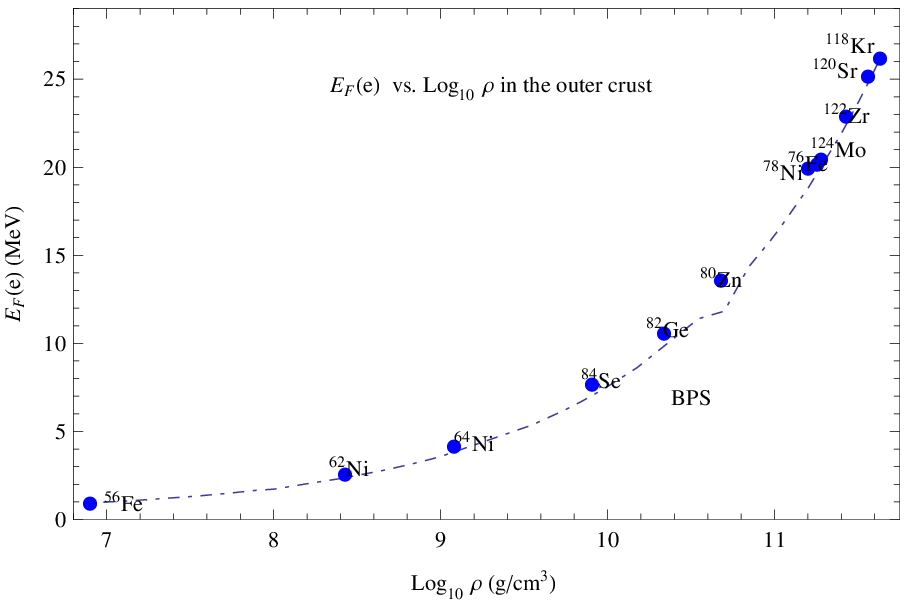}}\\
(a)&(b)\\
\end{tabular}
\end{center}
%\vspace*{8pt}
\caption{Numerically fitting in the outer crust. Left, the relation of $Y_{e}$
and $\rho$. Right, the relation of $E_{\rm F}(\rm e)$ and $\rho$. Dot-dashed
lines are the fitting curves, and circles denote the values of $Y_{e}$ (or
$E_{\rm F}(\rm e)$) and $\rho$ for each individual nuclide in BPS model.
 \label{fig1}}
\end{figure*}
Be note that the relation of $Y_{e}=Y_{p}= Z/A$ always
approximately holds in BPS model. From Table 2, it's  obvious that $Y_{e}$
decreases with $\rho$.  Combining Eq.(20) with the special solution, we plot the diagram of
$E_{\rm F}(\rm e)$ and $\rho$, and compare our results with those in BPS model.
From Fig. 1(b), it's obvious that $E_{\rm F}(e)$ increases with
$\rho$ in BPS model.  Theoretically, employing Eq.(20), we can obtain an approximate
value of $E_{\rm F}(e)$, given a matter density of the outer crust.

\subsection{Numerically fitting in the inner crust}
 The uncertainty of crust-core transition
density mainly comes from limited knowledge of EoS, especially the
density-dependence symmetry energy of neutron-rich nuclear matter
\cite{Lattimer01,Lattimer07}. Employing the method of quantum mechanics,
Negele \& Vautherin (1973)\cite{NV73}(hereinafter ``NV model'') firstly
obtained the value of $\rho_{t}\sim 1.32\times10^{14}$~g~cm$^{-3}$.
Recently, Atta \& Basu (2014) obtained the value of $1.54\times 10^{14}$~g~cm$^{-3}$
 (or$\rho_{t}\sim$ 0.0938~fm$^{-3}$) in
the $M3Y$ nucleon-nucleon effective
\begin{figure*}[htb]
\centerline{\psfig{file=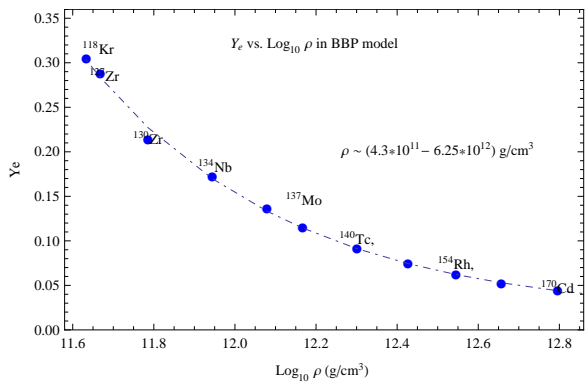,width=7.7cm}}
\vspace*{8pt}
\caption{The diagram of $Y_{e}$
vs. $\rm Log_{10}\rho$ in the lower density region in the inner crust.
The range of $\rho$ is (4.3$\times 10^{11}\sim 6.25\times 10^{12}$)
~g~cm$^{-3}$; Dot-dashed lines is the fitting curve, and circles
denote the values of $Y_{e}$ and $\rho$ for each individual nuclide in BBP model.
\protect \label{f2}}
\end{figure*}
interaction model\cite{Atta14}.
However, Atta \& Basu (2014)\cite{Atta14} obtained $Y_{e}\sim 0.031$,
which is very close to the value $Y_{e}\sim 0.032$ in NV model\cite{NV73}.

Here we select $\rho_{d}=4.3\times 10^{11}$~g~cm
$^{-3}$, corresponding to $Y_{e}\approx 0.3051$ and $E_{\rm F}
(e)\approx 26.2$~MeV, in BPS model, and select $\rho_{t}= 1.32
\times 10^{14}$~g~cm$^{-3}$ in NV model. Since the nuclei have
dissolved by merging together at the base of the inner crust,
the matter begins to turn into an ideal $npe$ system. From ST-83,
we obtain $Y_{e}\approx 0.0026$ and $E_{\rm F}(e)\approx 36.3$~MeV
at $\rho=\rho_{t}$.

Although the EoS above $\rho_{d}$ is reasonably well understood by BBP model,
this  model in a higher density region $\rho\geq 1.72\times
10^{14}$~g~cm$^{-3}$ was criticized by some authors\cite{Canuto74,Shapiro83,Gao13},
due to a monotonic and arbitrary increase of $Z$ with $A$, which causes
the relation of $P$ and $\rho$ to be changed
not much. Thus, we stop listing related calculations in the higher
density region. For convenience, the inner
crust can be roughly divided into two regions: a lower density region, and a
higher density region. The matter in the lower density region is still described by
BBP model, whereas matter in the higher density region is described by an
ideal mixed model(hereinafter ``IM model''). With respect to IM model, our
main hypotheses are as follows:

\begin{enumerate}
\item   Though matter in IM model is also a mixture of nuclei, free neutrons
     and free electron gas, the details of nucleon-nucleon interactions
     may be ignored, due to the uncertainty of nuclei compositions in IM model.
\item  The matter in IM model distributes in a range of $\sim 10^{12}\sim
      1.32\times 10^{14}$~g~cm$^{-3}$, $E_{\rm F}(e)$ grows with $\rho$
      smoothly, but the growth is not two much.
\item   Nuclei in IM model are more neutron-rich, and thus the values of $Y_{e}$ are
       universally smaller than those in BBP model, given the same matter densities.
\item  The electron fraction steadily decreases with matter density until $\rho=\rho_{t}$. To
      insure that $E_{\rm F}(e)$ increases with $\rho$, and $Y_{e}$ decrease with $\rho$ in IM model,
      the initial density of IM model, $\rho_{\rm IM}$, should be $1.0\times 10^{13}$~g~cm$^{-3}$, we arbitrarily select $\rho_{m}=6.25\times 10^{12}$~g~cm$^{-3}$ for ground state nucleus $^{170}_{48}Cd$ in BBP model to be $\rho_{IM}$ for IM model.
\end{enumerate}
Based on Table 1, we plot a diagram of $Y_{e}$ and $\rho$ in the lower density region of the inner crust, as shown in Fig. 2.
By numerically fitting, we obtain an analytical expression of
 $Y_{e}$ and $\rho$,
 \begin{equation}
 Y_{e}=0.0164-214.87e^{-\rm Log_{10}\rho}+3.67\times10^{9}e^{-2\rm Log_{10}\rho}.
 \label{21}
\end{equation}
in the range of 4.3$\times 10^{11}\sim 6.25\times 10^{12}$~g~cm$^{-3}$.
Due to lack of a detailed information on IM model introduced by this
work, we cannot give an exact formula of $Y_{e}$ and $\rho$ for
the higher density region. However, we mainly focus on the special
solution to $E_{\rm F}(e)$ and its potential applications. We assume
that $E_{\rm F}(e)$ grows with $\rho$ exponentially, and the expression of
 $E_{\rm F}(e)$ and $\rho$ has a simple exponential form,
 \begin{equation}
 E_{\rm F}(e)= A+ Be^{\rm Log_{10}\rho}~~{\rm MeV},
   \label{22}
\end{equation}
in the higher density region. Employing the boundary
conditions: (1) $\rho=\rho_{IM}=6.25\times 10^{12}$~g~cm$^{-3}$,
$Y_{e}\approx 0.0439$, $E_{\rm F}(e)\approx 33.43$~MeV, (2)
$\rho=\rho_{t}=1.32\times 10^{12}$~g~cm$^{-3}$, $Y_{e}\approx 0.0026$,
$E_{\rm F}(e)\approx 36.3$~MeV, we obtain the values
of two constants, $A=32.4$ and $B=2.882\times 10^{-6}$.  Combining the special
solution of Eq.(14) with Eq.(22), we obtain an analytical formula,
 \begin{equation}
 Y_{e}\approx0.005647(\frac{\rho_{0}}{\rho})(0.539+4.803\times10^{-8}e^{\rm Log_{10}\rho})^{3},
  \label{23}
\end{equation}
in the higher density region.
The relations of $Y_{e}$ and $\rho$, and $E_{\rm F}(e)$ and
$\rho$ in the inner crust are shown in Fig.3.
\begin{figure}[pb]
\begin{center}
\begin{tabular}{cc}
\scalebox{0.65}{\includegraphics{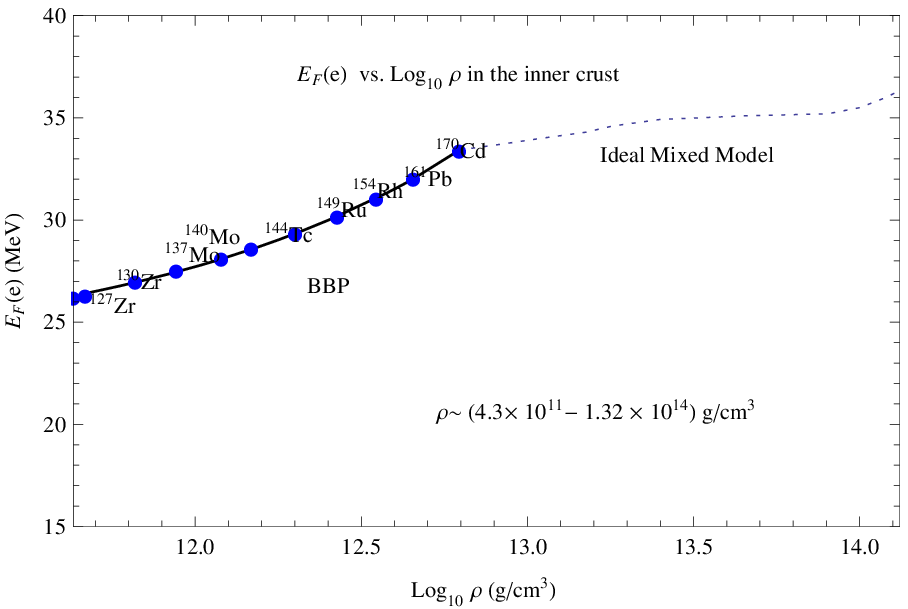}}&\scalebox{0.65}{\includegraphics{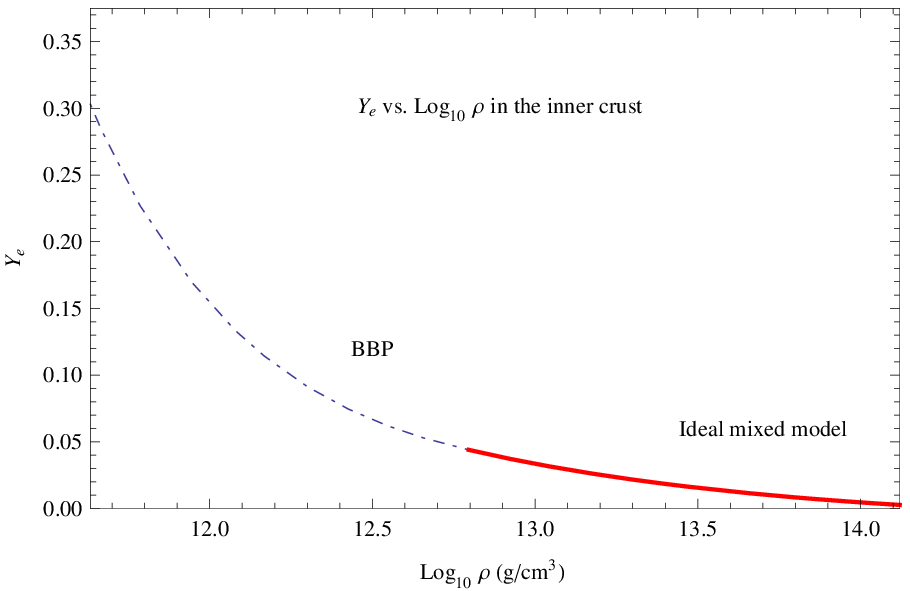}}\\
(a)&(b)\\
\end{tabular}
\end{center}
%\vspace*{8pt}
\caption{Numerically fitting in the inner crust.
 Left, the relation of $Y_{e}$ and $\rho$. Right, the relation of $E_{\rm F}(\rm e)$
 and $\rho$. The range of $\rho$ is (4.3$\times 10^{11}\sim 1.32\times 10^{14}$)
 ~g~cm$^{-3}$; Dot-dashed lines are for BBP model; and dotted line and solid line are for IM model.
 \label{fig3}}
\end{figure}

In addition, we calculate the values of $E_{\rm F}(e)$ and
$Y_{e}$ in IM model , as listed in column 4 and column 5 of Table 3, respectively.
\begin{table*}[htb]
\tbl{Values of $E_{\rm F}(e)$ and $Y_{e}$ in BBP model
and IM model.}
{\begin{tabular}{@{}cccccccccc@{}} \toprule
$\rho$  &$Y_{e}^{'}$ & $E_{\rm F}(e)^{'}$& $E_{\rm F}(e)$ &$Y_{e}$&$\rho$  &$Y_{e}^{'}$ & $E_{\rm F}(e)^{'}$& $E_{\rm F}(e)$ &$Y_{e}$ \\
(g~cm$^{-3}$)    &        &MeV & MeV & & (g~cm$^{-3}$)    &        &MeV & MeV & \\
\colrule
6.25$\times 10^{12}$ &0.0439 &33.43 &33.43 &0.0439 & 3.44$\times 10^{13}$ &0.0236 &48.10 &34.57 &0.0088\\
8.38$\times 10^{12}$ &0.0371 &34.98 &33.57 &0.0434  & 4.68$\times 10^{13}$ &0.0233 &52.95 &34.88 &0.0066\\
1.10$\times 10^{13}$ &0.0326 &36.68 &33.72 &0.0255 & 5.96$\times 10^{13}$ &0.0235 &57.56 &35.16 &0.0053 \\
1.50$\times 10^{13}$ &0.0289 &39.00 &33.91 &0.0190 & 8.01$\times 10^{13}$ &0.0242 &64.32 &35.54 &0.0041 \\
1.99$\times 10^{13}$ &0.0264 &41.58 &34.11 &0.0146 & 1.99$\times 10^{13}$ &0.0264 &41.58 &34.11 &0.0146\\
2.58$\times 10^{13}$ &0.0246 &44.37 &34.31 &0.0115 & 2.58$\times 10^{13}$ &0.0246 &44.37 &34.31 &0.0115\\
 \botrule
\end{tabular} \label{ta3}}
\end{table*}
The data of columns 1, 2, 3, 6, 7 and 8 are
cited from Table 1 in this paper. The data of
columns 4, 5, 9 and 10 are obtained from IM model using
Eq.(14) and Eq.(24). From Table 3, IM model may be superior
to BBP model in the higher density region.

\subsection{Numerically fitting in the outer core}
The numerical simulations in the outer core are usually subject
to two distinct uncertainties:(1) determining the nuclear potential
for nucleon-nucleon interaction, and (2) finding an appropriate
technique for solving the many-body problem. The nuclear potential
is constrained somewhat by nucleon-nucleon scattering data and
nuclear matter results\cite{Lattimer91,Haensel07}.

As mentioned in Section 3.2, the matter in the outer core
contains relativistic electrons, non-relativistic
neutrons, and non-relativistic protons. Here, for the sake of simplicity, we consider
a homogenous ideal $npe$ system under $\beta$-equilibrium, and
adopt ST-83 approximation\cite{Shapiro83}) as
the main method to treat EoS of this
system. The relation of $Y_{e}$ and $\rho$ is described by
\begin{equation}
Y_{e}\approx \frac{n_{e}}{n_{n}}= 0.005647\times(\frac{\rho}{\rho_{0}})~,
\label{24}
\end{equation}
in the density range of $0.5\rho_{0}\sim 2.5\rho_{0}$ (c.f. Sec. 2.2),
Combining the above equation with the special solution, we plot schematic
diagrams of $Y_{e}$ vs. $\rm Log_{10}\rho$ and $E_{\rm F}(e)$ vs. $\rm Log_{10}\rho$
in the outer core of a NS, as shown in Fig. 4.
\begin{figure*}[htb]
\begin{center}
\begin{tabular}{cc}
\scalebox{0.65}{\includegraphics{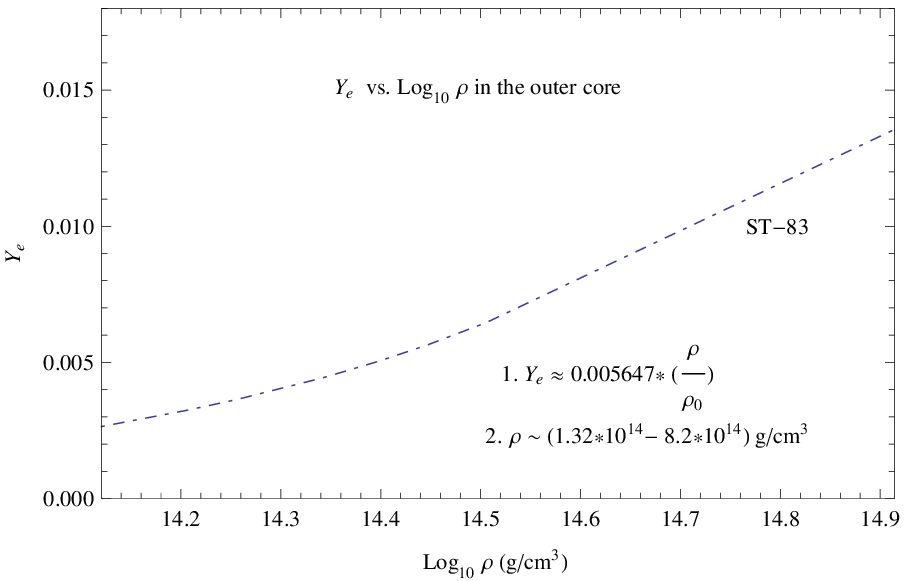}}&\scalebox{0.65}{\includegraphics{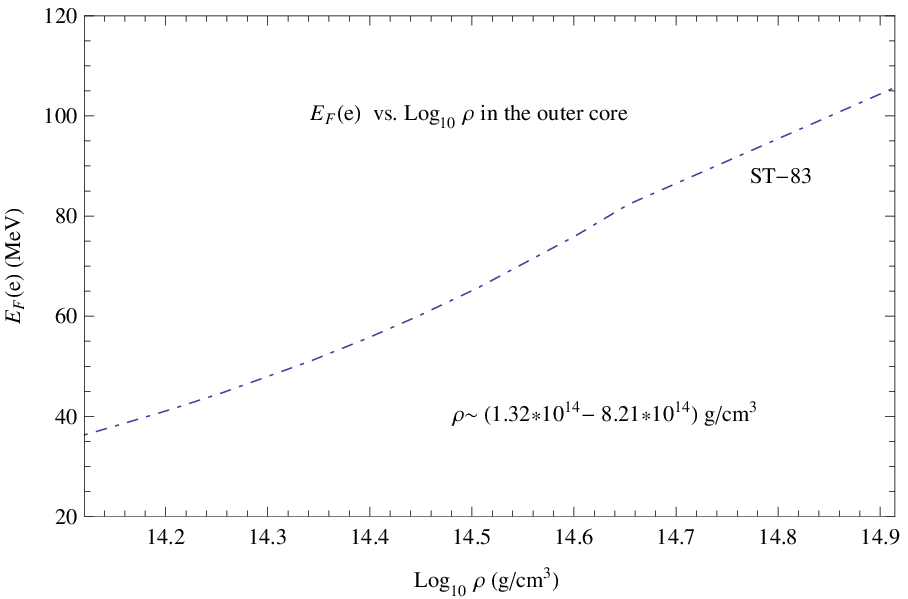}}\\
(a)&(b)\\
\end{tabular}
\end{center}
%\vspace*{8pt}
\caption{Numerically fitting in the outer core.
 Left, the relation of $Y_{e}$ and $\rho$. Right, the relation of $E_{\rm F}(e)$
 and $\rho$. The range of $\rho$ is (1.32$\times 10^{14}\sim 8.2\times 10^{14}$)
 ~g~cm$^{-3}$.
 \label{fig4}}
\end{figure*}

\subsection{Numerically simulating in the inner core}
Although the maximum
inner-core-density could exceed the transition density
$\rho_{tr}$\cite{Tsuruta09}, for the sake of simplicity, we focus on a
non-relativistic domain of $\sim 2.5\rho_{0}\sim 2.0\times 10^{15}$~g~cm$^{-3}$,
and consider a system, composed of neutrons, protons, electrons and muons under
$\beta$-equilibrium. Here our first aim is to deduce an analytical expression of
$Y_{e}$ and $\rho$ in the inner core. A detailed deduction is summarized as follows:

When $E_{\rm F}(e)$ is high enough, it is energetically
favorable for electrons to turn into muons, so that muons and electrons are in
equilibrium: $\mu^{-}\leftrightarrow e^{-}$, here we have as usual assumed
that the neutrinos leave the system. Once we know this equilibrium, thermodynamics
does not require us to know any detail of the process of $\mu^{-}\leftrightarrow e^{-}$.
Chemical potential equilibrium and charge neutrality give
\begin{eqnarray}
&&m_{\mu}c^{2}(1+x_{\mu}^{2})^{1/2}= m_{e}c^{2}(1+x_{e}^{2})^{1/2} ,\nonumber\\
&&m_{n}c^{2}(1+x_{n}^{2})^{1/2}= m_{p}c^{2}(1+x_{p}^{2})^{1/2}+ m_{e}c^{2}(1+x_{e}^{2})^{1/2},~~\nonumber\\
&&(m_{p}x_{p})^{3}=(m_{e}x_{e})^{3}+(m_{\mu}x_{\mu})^{3}~.
\label{24}~~
\end{eqnarray}
where $x_{i}=p_{\rm F}(i)/m_{i}c^{2}$ ($i= n, p, e£¬\mu$) is the dimensionless Fermi momenta.
From Eq.(24), the electron fraction can be expressed as a function
of $x_{e}$,
\begin{eqnarray}
&&Y_{e}=\frac{n_{e}}{n_{B}}\approx \frac{n_{e}}{n_{n}}
=\frac{(m_{e}x_{e})^{3}}{(m_{n}x_{n})^{3}}=\frac{m_{e}^{3}x_{e}^{3}}{m_{n}^{3}\{[\frac{m_{p}}{m_{n}}(1+x_{p}^{2})^{\frac{1}{2}}+
\frac{m_{e}}{m_{n}}(1+x_{e}^{2})^{\frac{1}{2}}]^{2}-1 \}^{\frac{3}{2}}}~,\nonumber\\
&& =\frac{m_{e}^{3}x_{e}^{3}}{m_{n}^{3}\{[1+\frac{m_{e}^{3}x_{e}^{3}+
m_{\mu}^{3}(\frac{m_{e}^{2}x_{e}^{2}}{m_{\mu}^{2}}-1)^{\frac{3}{2}}}{2m_{p}^{2}}+
\frac{m_{e}(1+x_{e}^{2})^{\frac{1}{2}}}{m_{n}}]^{2}-1\}^{\frac{3}{2}}}~~,
\label{25}~
\end{eqnarray}
where we used $m_{p}/m_{n}\approx 1$ and $\mu_{\mu}=\mu_{e}$. Inserting
$m_{e}=0.511$~MeV, $m_{p}=938.28$~MeV, $m_{n}=939.57$~MeV,
and $m_{\mu}=105.7$~MeV, into Eqs.(25), we obtain a schematic diagram of $Y_{e}$ and $x_{e}$ in the inner core of a NS.
\begin{figure*}[htb]
\centerline{\psfig{file=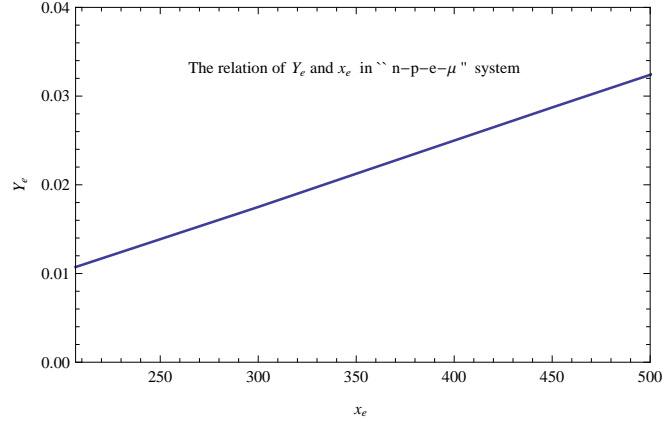,width=8.7cm}}
\vspace*{8pt}
\caption{The relation of $Y_{e}$ and $x_{e}$ for $npe-\mu$ system
 in the inner core of a NS.
\protect \label{f5}}
\end{figure*}
From Fig. 5, it's obvious that $Y_{e}$ increases
with $x_{e}$ in the inner core of a NS.

The threshold density for muons to appear, $\rho_{\mu}$, is also an important issue.
When at $\rho_{\mu}$, $n_{\mu}=0$ that is $x_{\mu}=0$.  Since the
electrons are highly relativistic, we can take $x_{e}\gg 1$, Eq.(24) becomes
\begin{eqnarray}
&& m_{\mu}= m_{e}x_{e}~,\nonumber\\
&&m_{n}c^{2}(1+x_{n}^{2})^{\frac{1}{2}}=m_{p}c^{2}(1+x_{p}^{2})^{\frac{1}{2}}+ m_{e}x_{e},\nonumber\\
&&m_{p}x_{p}= m_{e}x_{e}. ~\label{26}~
\end{eqnarray}
Inserting the values of $m_{e}$, $m_{p}$, $m_{n}$, and $m_{\mu}$ into Eq.(26),
we get $x_{e}=206.8$,$x_{p}=0.1126$, and $x_{n}=0.4986$, corresponding to
$\rho_{\mu}=8.21\times10^{14}$~g~cm$^{-3}$. Combining Eq.(25) with the expression of
$x_{e}=1.0088\times10^{-2}(Y_{e}\rho)^{\frac{1}{3}}$~ with , we
calculate the values of $Y_{e}$ in the range of $8.2\times 10^{14}\sim
2.0\times 10^{15}$~g~cm$^{-3}$, as shown in Fig. 6(a). By fitting these values
of $Y_{e}$ and $\rho$, we obtain an analytical formula of
$Y_{e}$ and $\rho$
\begin{equation}
Y_{e}\approx -0.0115+ 7.402\times 10^{-9}e^{\rm Log_{10}\rho},
\label{27}~
\end{equation}
in the inner core of a NS. Combining Eq.(27) with the special
solution, we plot the diagram of $E_{\rm F}(e)$ and $\rho$, as shown
in Fig. 6(b).
\begin{figure*}[htb]
\begin{center}
\begin{tabular}{cc}
\scalebox{0.69}{\includegraphics{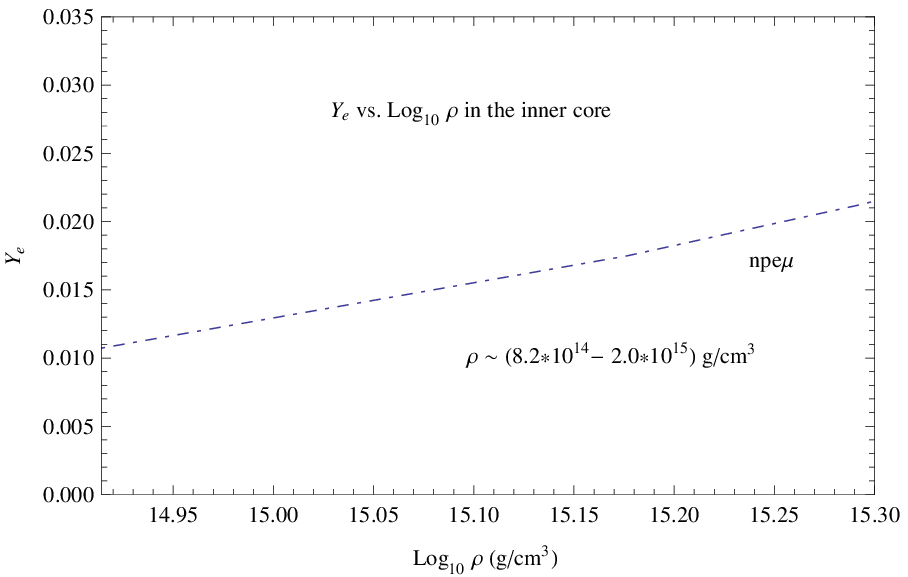}}&\scalebox{0.69}{\includegraphics{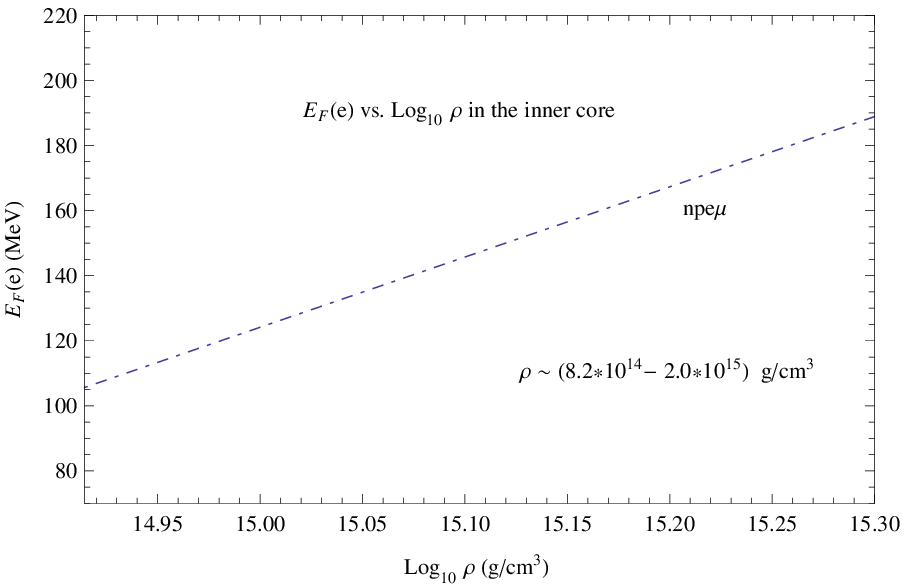}}\\
(a)&(b)\\
\end{tabular}
\end{center}
%\vspace*{8pt}
\caption{Numerically fitting in the inner core.
 Left, the relation of $Y_{e}$ and $\rho$. Right, the
 relation of $E_{\rm F}(\rm e)$ and $\rho$. The range of $\rho$ is
 assumed to be 8.2$\times 10^{14}\sim 2.0\times 10^{15}$~g~cm$^{-3}$ arbitrarily.
 \label{fig6}}
\end{figure*}
Here we arbitrarily select a maximum density $2.0\times10^{15}$~g~cm$^{-3}$
for the inner core matter, because at
higher densities (more than $10^{15}$~g~cm$^{-3}$), the
composition is expected to include an appreciable number of hyperons,
and the nucleon interactions must be treated relativistically.
In a real scenario, matter in the inner core could be more complicated
than that of ideal $npe\mu$ system.

\subsection{Summary}
As to the main purpose and innovations for the simulations above, a brief summary is presented as follows:
\begin{figure*}[htb]
\begin{center}
\begin{tabular}{cc}
\scalebox{0.65}{\includegraphics{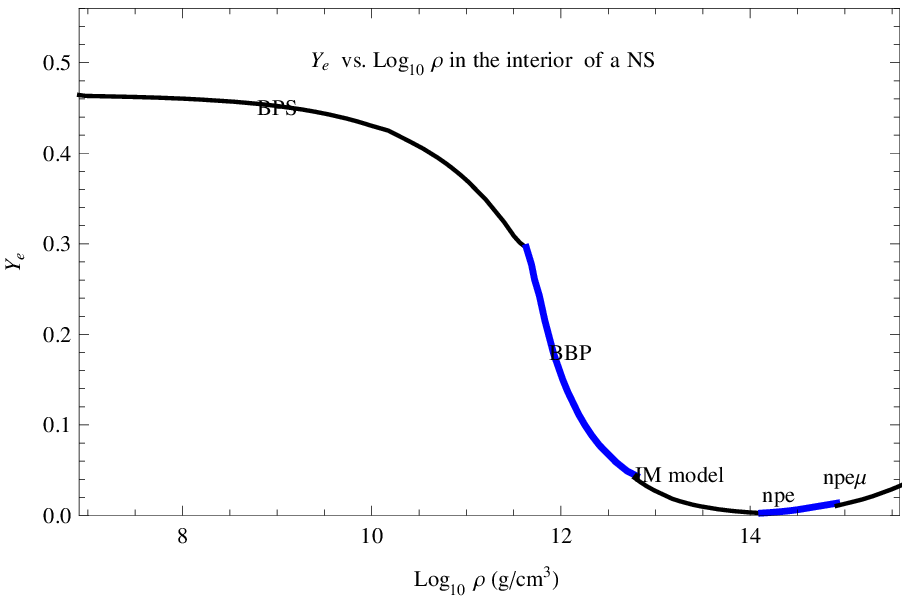}}&\scalebox{0.65}{\includegraphics{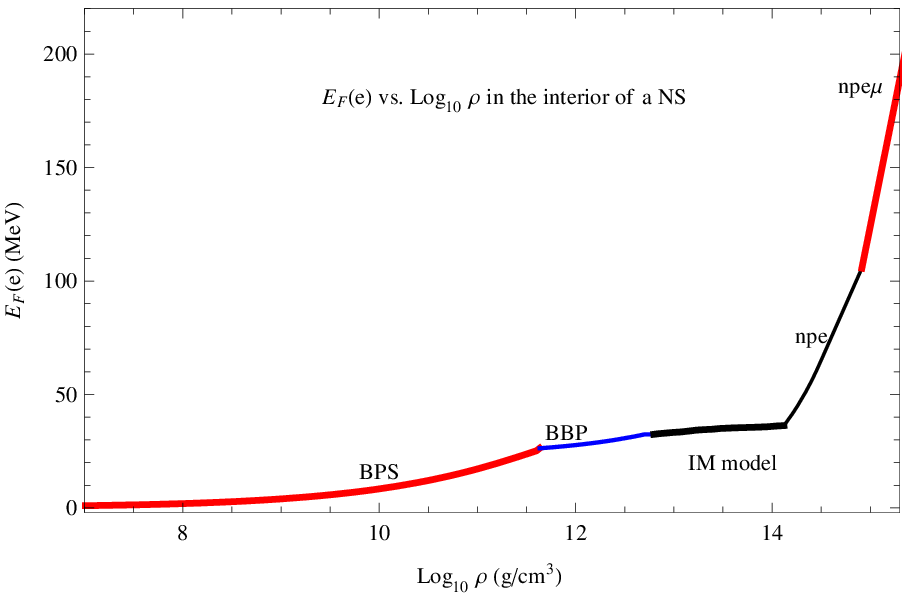}}\\
(a)&(b)\\
\end{tabular}
\end{center}
%\vspace*{8pt}
\caption{Numerically fitting $Y_{e}$ and $E_{\rm F}(e)$ in the whole interior of a NS.
 Left, the relation of $Y_{e}$ and $\rho$. Right, the relation of $E_{\rm F}(e)$
 and $\rho$.
 \label{fig7}}
\end{figure*}
\begin{enumerate}
\item Firstly, in the previous studies, there are too many matter models, some of
which are rather successful and thus representative. By solving the EoSs of these models
(including the above models), we can obtain the relations of $E_{\rm F}(e)$ vs. $\rho$
and $Y_{e}$ vs. $\rho$.  However, there still exist limitations for each EoS to
some extent. For example, any EoS has a certain application range of matter density, the
resulting values of $E_{\rm F}(e)$ and $Y_{e}$ will deviate from those in the actual scenario
if the EoS is beyond of its application range of density.

\item  Secondly, up to now, we have not obtained a uniform EoS of one certain
matter model, which is excellent in describing the relations of $E_{\rm F}(e)$,
$Y_{e}$ and $\rho$ in the whole interior of a NS. By numerically simulating, we
have obtained a set of analytical expressions for $E_{\rm F}(e)$, $Y_{e}$ and $\rho$
from EoSs. However, this is not our main purpose. Our ultimate aim is to investigate
the whole variation trends in $Y_{e}$ vs. $\rho$ and $E_{\rm F}(e)$  vs. $\rho$ in the
whole relativistic electron matter region of a NS using these analytical expressions obtained.

\item Thirdly, as we know, the electron Fermi energy, as well as the
electron pressure, is always continuous in the interior
of a NS. From the general formula for $E_{\rm F}(e)$ (see Eq.(10)),
$Y_{e}$ is also a continuous function of $\rho$ in a NS. Utilizing the boundary conditions and the
fitting formula of $E_{\rm F}(e)$, $Y_{e}$ and $\rho$ in four matter regions (the outer
crust, inner crust, outer core, and inner core), we obtain the values of $E_{\rm F}(e)$
and $Y_{e}$, which are smooth and continuous functions of matter density, as well shown in Fig.7.

\item At last, we can see that the electron fraction $Y_{e}$ firstly decreases in the crust, then increases in the core (from Fig.7(a)). The decrease in $Y_{e}$ is caused by increasing neutronization of nuclei, as the depths of the crust increases, whereas the increase in $Y_{e}$ is due to the fact that, the proton fraction $Y_p$ always increases with the core matter density required by the chemical potential equilibrium under $\beta-$ equilibrium, and the charge neutrality gives $n_p=n_e + n_{\mu}$. Be note that $n_{\mu}$ is far less than $n_e$ given a matter density. From Fig.7(b), the electron Fermi energy always increases with matter density, the reason for the increase in $E_{\rm F}(e)$ is that $E_{\rm F}(e)$ solely depends on $n_e$, which always increases with $\rho$ (for details see in Sec.6.)
\end{enumerate}
It should be admitted that our method of fitting $Y_{e}$ and $E_{\rm F}(e)$ is somewhat
simple, but is practically convenient. Especially, it is convenient to
calculate the value of $E_{\rm F}(e)$ given a matter density inside a NS.
\section{Numerically Simulating $E_{\rm F}(e)$ and $Ye$ in A NS (II)}\label{V}
As mentioned in Sec.4, numerically fitting $E_{\rm F}(e)$ and $Y_{e}$
in the core of a NS are subject to many
uncertainties, including the possibility of neutron and proton
superfluid, of pion condensation, of phase transitions to quark
matter\cite{Du09,Lai13}, and the consequences
of the $\Delta$ resonances\cite{Xu13a,Xu13b}.  To date, many of relativistic
models\cite{Glendenning85,Schaffner93,Shen02,Yang08,Dutra12}
have drawn attentions in investigating EoSs  because they are
particularly suited for describing NSs according to the special
relativity. The most common among them is the relativistic-mean-
field (RMF) theory, which has become a standard method to study
nuclear matter and finite-nuclei properties.

As an up-to-date representative theoretical work in RMF theory is Dutra et al. (2014)\cite{Dutra14},
in which 263 RMF models are examined three different sets of constraints related to
pure neutron matter, symmetric nuclear matter, symmetry energy, and its derivatives are used.
The authors gave a detailed discussion of these models when investigating properties like realistic values of the asymmetry energy and its slope, compressibility and the question that whether the EoS can actually reproduce 2-solar mass stars, as have been observed in PSR J1615-2230\cite{Demorest10}, within the whole NS interior, and added extensive supernova EOS online tables (like on http://compose.obspm.fr).

According to RMF theory, baryonic interactions include three cases: (1) exchanging
three mesons of $\sigma$, $\omega$ and $\rho$; (2) exchanging four mesons of $\sigma$, $\omega$,
$\rho$ and $\sigma$; (3) exchanging five mesons of $\sigma$, $\omega$,
$\rho$, $\sigma^{*}$, $\phi$. In order to provide a new insight into application of
RMF models to the nuclear matter properties, Dutra et al. (2014) considered the four-meson-exchange
baryonic interactions, and divided the RMF models into seven types regarding their lagrangian density structures:
Typ1-1: linear finite range models in which
$A=B=C=\alpha_{1}=\alpha_{2}=\alpha^{'}_{1}=\alpha^{'}_{2}=\alpha^{'}_{3}=g_{\sigma}=0$; Type-2:$\sigma^{3}+\sigma^{4}$ models in which
$C=\alpha_{1}=\alpha_{2}=\alpha^{'}_{1}=\alpha^{'}_{2}=\alpha^{'}_{3}=g_{\sigma}=0$ ; Type-3: $\sigma^{3}+\sigma^{4}+\omega^{4}_{0}$
models in which $\alpha_{1}=\alpha_{2}=\alpha^{'}_{1}=\alpha^{'}_{2}=\alpha^{'}_{3}=g_{\sigma}=0$; Type-4: $\sigma^{3}+\sigma^{4}+\omega^{4}_{0}+$
cross terms models in which $g_{\sigma}=0$ and at least one of the coupling constants, $\alpha_{1}, \alpha_{2}, \alpha^{'}_{1}, \alpha^{'}_{2}$, or
$\alpha^{'}_{3}$ is different from zero; Type-5: density-dependent models in which $g_\delta \rightarrow T_{\delta}(\rho)=0$; Type-6: point-coupling models in which $g_\delta \neq 0$; and Type-7: $\delta-$ meson models in which $a_{TS}=0$. For the physical meaning of each parameter above, see Dutra et al. (2014).

In this paper, for the purpose of comparing with the classical models applied in Sec.4, we will choose
Type-2 in Dutra et al. (2014) as a representative RMF theoretical work. The main merit of Type-2 in Dutra et al (2014) lies in that it describes well the properties of EoS of a NS in the vicinity of the saturated nuclear density $\rho_{0}$.

From Eq.(1) of Dutra et al. (2014), we get the effective lagrangian of Type-2 \textbf{in Dutra et al (2014)}
\begin{eqnarray}
&& \mathcal{L}=\sum_{B}\bar{\psi}_{B}\left[ i\gamma _{\mu }\partial
^{\mu }-m_{B}-g_{\sigma B}\sigma -g_{\omega B}\gamma _{\mu }\omega ^{\mu }
-g_{\rho B}\gamma _{\mu }\tau_{i}\rho _{i}^{\mu }\right] \psi _{B} \nonumber \\
&&+\frac{1}{2}\partial _{\mu }\sigma
\partial ^{\mu }\sigma -\frac{1}{2}m_{\sigma }^{2}\sigma ^{2}
-\frac{1}{3}g_{2}\sigma ^{3}-\frac{1}{4}g_{3}\sigma ^{4}-\frac{1}{4}W_{\mu
\nu }W^{\mu \nu }+\frac{1}{2}m_{\omega }^{2}\omega _{\mu }\omega ^{\mu }\nonumber \\
&&-\frac{1}{4}R_{i\mu \nu }R_{i}^{\mu \nu }+\frac{1}{2}m_{\rho }^{2}\rho _{i\mu }\rho
_{i}^{\mu }+\sum_{l}\bar{\psi}_{l}\left[ i\gamma _{\mu }\partial ^{\mu }-m_{l}\right]
\psi _{l},  \label{29}
\end{eqnarray}%
where the baryon species are marked as B,and the sum on $l$ is over electrons and muons ($e^{-}$ and $\mu
^{-}$). In the RMF models, the meson fields are treated as classical
fields, and the field operators are replaced by their expectation
values. The meson field equations in uniform matter have the
following form:
\begin{eqnarray}
&&m_{\sigma }^{2}\sigma +g_{2}\sigma ^{2}+g_{3}\sigma ^{3}=-\sum_{B}\frac{%
g_{\sigma B}}{\pi ^{2}}\int_{0}^{k_{F}^{B}}\frac{m_{B}^{\ast }}{\sqrt{%
k^{2}+m_{B}^{\ast 2}}}k^{2}dk,  \label{30} \\
&&m_{\omega }^{2}\omega +c_{3}\omega ^{3}=\sum_{B}\frac{g_{\omega B}\left(
k_{F}^{B}\right) ^{3}}{3\pi ^{2}},  \label{31} \\
&&m_{\rho }^{2}\rho =\sum_{B}\frac{g_{\rho B}\tau _{3B}\left(
k_{F}^{B}\right) ^{3}}{3\pi ^{2}},  \label{32}
\end{eqnarray}%
where $\sigma =\left\langle \sigma \right\rangle ,$ $\omega =\left\langle
\omega^{0}\right\rangle ,$ and $\rho =\left\langle \rho ^{30}\right\rangle ,$
are the nonvanishing expectation values of meson fields in NS matter,
$m_{B}^{\ast}=m_{B}+g_{\sigma B}\sigma $ is the effective mass of the baryon species $B$, and $k_{F}^{B}$
is the Fermi momentum. At zero temperature the lepton chemical potentials are expressed by
\begin{equation}
\mu _{l}=\sqrt{{k_{F}^{l}}^{2}+m_{l}^{2}}, ~~~~(\rm fm^{-1}) \label{33}
\end{equation}
The charge neutrality condition is given by
\begin{equation}
\sum_{B}q_{B}\rho_{B}-n_{e}-n_{\mu}=0,  \label{34}
\end{equation}
where $\sum_{B}q_{B}=n_{B}$, and $q_{B}$ is the baryon electric charge.
We can solve the coupled equations self-consistently at a
given baryon density. Then we get the total
energy density $\varepsilon$ and pressure $P$
\begin{eqnarray}
&&\varepsilon =\sum_{B}\frac{1}{\pi ^{2}}\int_{0}^{k_{F}^{B}}\sqrt{%
k^{2}+m_{B}^{\ast 2}}\ k^{2}dk+\frac{1}{2}m_{\sigma }^{2}\sigma ^{2}+\frac{1%
}{3}g_{2}\sigma ^{3}+\frac{1}{4}g_{3}\sigma^{4}  \nonumber \\
&&+\frac{1}{2}m_{\omega}^{2}\omega ^{2}+\frac{3}{4}c_{3}\omega ^{4}+\frac{1%
}{2}m_{\rho }^{2}\rho^{2}+\sum_{l}\frac{1}{\pi ^{2}}\int_{0}^{k_{F}^{l}}\sqrt{k^{2}+m_{l}^{2}}\
k^{2}dk,  \label{35}
\end{eqnarray}%
\begin{eqnarray}
&& P=\frac{1}{3}\sum_{B}\frac{1}{\pi ^{2}}\int_{0}^{k_{F}^{B}}\frac{%
k^{4}\ dk}{\sqrt{k^{2}+m_{B}^{\ast 2}}}-\frac{1}{2}m_{\sigma }^{2}\sigma
^{2}-\frac{1}{3}g_{2}\sigma ^{3}-\frac{1}{4}g_{3}\sigma^{4}  \nonumber \\
&&+\frac{1}{2}m_{\omega }^{2}\omega ^{2}+\frac{1}{4}c_{3}\omega^{4}+\frac{1%
}{2}m_{\rho }^{2}\rho ^{2}+\frac{1}{3}\sum_{l}\frac{1}{\pi ^{2}}\int_{0}^{k_{F}^{l}}
\frac{k^{4}\ dk}{\sqrt{k^{2}+m_{l}^{2}}}.  \label{36}
\end{eqnarray}

In the work of Dutra et al. (2014), the authors
listed more than 130 RMF models for Type-2. These models have been
widely used because of several important aspects not always
present in non-relativistic models (e.g., intrinsic
Lorentz covariance, appropriate saturation mechanism for
nuclear matter and etc). Unfortunately, the authors didn't present any information on
the electron Fermi energy and electron fraction within these models,
due to the restriction of space.

In order to obtain better agreement
for the behavior of EOS at high matter densities with that predicted
by the relativistic Brueckner-Hartree-Fock theory, the parameter set TMA
was developed, based on two other widely used
parameter sets, TM1 and TM2\cite{Sugahara94}. The parameter set TMA has
been one of the most successful modern parameter sets.
In this work, we select the parameter set TMA \cite{Geng05,Singh12} for the mean-field Lagrangian density,
and list the parameter values in Table 4.

\begin{table*}[htb]
\tbl{The parameter values of the effective force TMA used
in the calculation.}
{\begin{tabular}{@{}cccccccccc@{}} \toprule
$m_N$ & $m_\sigma$ & $m_\omega$ & $m_\rho$ & $g_{\sigma}$& $g_{\omega}$ & $g_{\rho}$ & $g_2$  & $g_3$ &$c_3$  \\
MeV & MeV & MeV &MeV  & (fm)$^{-1}$& (fm)$^{-1}$ & (fm)$^{-1}$&(fm)$^{-1}$ & (fm)$^{-1}$&(fm)$^{-1}$ \\
\colrule
 939.0 & 519.151 & 781.950 & 768.100 & 10.055 & 12.842 & 3.800 & -0.328 & 38.862 & 151.590\\
\botrule
\end{tabular} \label{ta4}}
\end{table*}
In addition, the properties of ground-state nuclear matter include: the saturation density $\rho_0$=0.147\,(fm)$^{-3}$, bulk binding energy/nucleon $(E/A)_\infty=-16.0$\,MeV, incompressibility
$K=318.0$~MeV, bulk symmetry energy/nucleon $a_{sym}=30.68$\,MeV, and the effective mass ratio $m^{*}/m=0.635$\cite{Geng05,Singh12}.

Inserting the above parameters into Eq.(34) and Eq.(35), we get the relation of
$\varepsilon$ and $P$ in Dutra et al.(2014) (Type-2), shown as in Fig.8.
\begin{figure*}[htb]
\centerline{\psfig{file=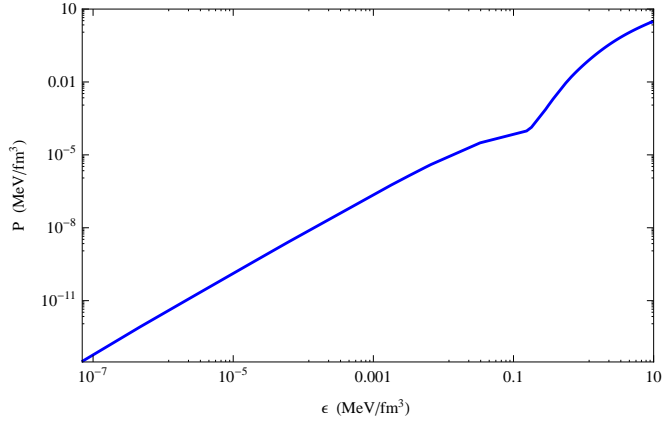,width=8.7cm}}
\vspace*{8pt}
\caption{The relation of $\varepsilon$ and $P$ for RMFT in Dutra et al.(2014)(Type-2) .
\protect \label{f8}}
\end{figure*}
Here the relation of $c\hbar$=1 MeV$\times 5.43\times 10^{-3}$~fm$^{-1}$ is used.
From Fig. 8, it's obvious that $P$ increases
with $\varepsilon$ in the whole interior of a NS.
Inserting EoSs (Eq.(34) and Eq.(35)) into TOV equation
\begin{equation}
\frac{dP}{dr}=-\frac{G}{r} \frac{\left(\varepsilon+P\right)\left(m+ 4\pi r^{2}P\right)}{r-2Gm},~~~~~
\frac{dm}{dr}= 4\pi r^{2}\varepsilon,  \label{36}
\end{equation}
gives the one-to-one relation of star mass $m$ and radius $r$.
Since the baron number density at the stellar center is determined
by $n_{B}(c)=\frac{m}{m_{u}}/(\frac{4}{3}\pi r^{3})$, we obtain the relation
of star mass $m$ and $n_{B}(c)$, as shown in Fig.9.
\begin{figure*}[pb]
\centerline{\psfig{file=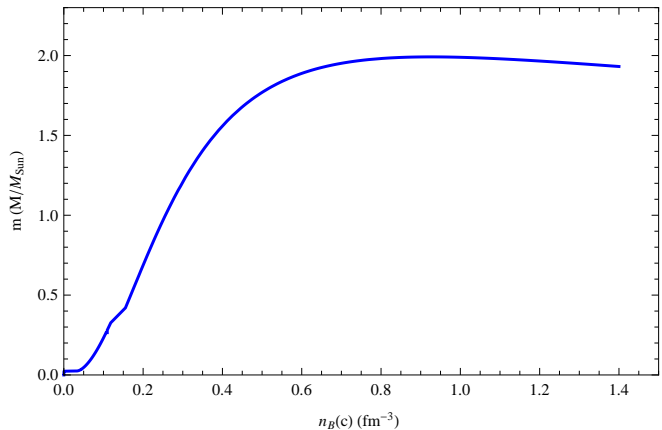,width=8.7cm}}
\vspace*{8pt}
\caption{The relation of $m$ and $n_{B}(c)$ in Dutra et al. (2014)(Type-2).
\protect \label{f9}}
\end{figure*}
From Fig.9, it's obvious that $m$ increases
with the baron number density at the stellar center $n_{B}(c)$,
and the maximum stellar mass is about 1.9916\,$M_{Sun}$
($M_{Sun}$ is the solar mass), corresponding
to $n_{B}(c)\textbf{=0.9156}$ fm$^{-3}$ in Dutra et al. (2014)(Type-2) for RMF theory.
The maximum stellar mass of 1.9916\,$M_{Sun}$) is very close to the observational NS mass limit of
 $\sim 1.97M_{Sun}$ for PSR J1615-2230\cite{Demorest10}, which indicates our parameter-set
 choice (TMA) is rather reliable and successful.

We also calculate the values of $E_{\rm F}(e)$ and $Y_{e}$
in the whole density range in Dutra et al. (2014)(Type-2),
as listed in Appendix B.  Based on Tables 5-8 in Appendix B,
we numerically fit the relations of $Y_e$ vs. $Log_{10}\rho$ and
$E_{\rm F}(e)$ vs. $Log_{10}\rho$ in Dutra et al. (2014)
(Type-2), as shown in Fig.10, and obtain a set of analytical
expressions of
\begin{eqnarray}
&&Y_{e}=-0.000012+1.213\times10^{-9}e^{\rm Log_{10}\rho}-2.51\times10^{-15}e^{2\rm Log_{10}\rho}
+8.19\times10^{-21}e^{3\rm Log_{10}\rho}, \nonumber\\
&&Y_{e}=0.0644-1.506\times10^{-7}e^{\rm Log_{10}\rho}+1.11\times10^{-13}e^{2\rm Log_{10}\rho}
+1.84\times10^{-20}e^{3Log_{10}\rho},        \nonumber\\
&&Y_{e}=-0.188+1.85\times10^{-7}e^{\rm Log_{10}\rho}-3.59\times10^{-14}e^{2\rm Log_{10}\rho}
+2.54\times10^{-21}e^{3\rm Log_{10}\rho},
\label{37}~
\end{eqnarray}
for $\rho\sim 6.92\times 10^{11}-6.07\times 10^{13}$, $6.07\times 10^{13}-6.9\times 10^{14}$,
and $6.9\times 10^{14}-2.56\times 10^{15}$~g~cm$^{-3}$, respectively.

Combining Eq.(37) with the special solution of Eq.(14), we plot the diagram of
$E_{\rm F}(e)$ vs. $Log_{10}\rho$ (see the solid-line of Fig.10(b)),  and compare our results
with those in Dutra et al.(2014)(Type-2)(see the dotted-line of Fig.10(b)). As shown in Fig.10(b),
the fitted values of electron Fermi energies are well in agreement with their calculated values, which
indicates that the pairs of $E_{\rm F}(e)$ and $Y_e$ in Tables 5-8 in Appendix B are suited for
the special solution of Eq.(14). In other words, the special solution of Eq.(14) is equally applicable to
RMF models.

We have found that both $Y_{e}$ and $E_{\rm F}(e)$ increase with
matter density in Dutra et al. (2014)(Type-2). Here, the maximum of
central density is arbitrarily selected to be $\rho=$1.4 fm$^{-3}$, corresponding to
$Y_{e}=0.15570$ and $E_{\rm F}(e)=367.40$ MeV. Like in classical models, the increase in
$E_{\rm F}(e)$ in RMF theory is also caused by an increase in electron number density in the core of a NS.
 \begin{figure*}[htb]
\begin{center}
\begin{tabular}{cc}
\scalebox{0.65}{\includegraphics{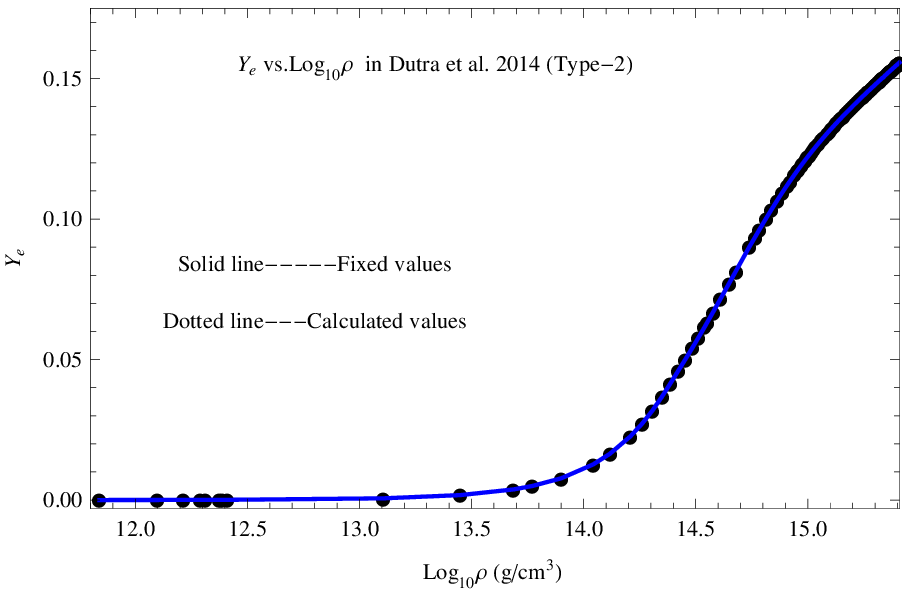}}&\scalebox{0.65}{\includegraphics{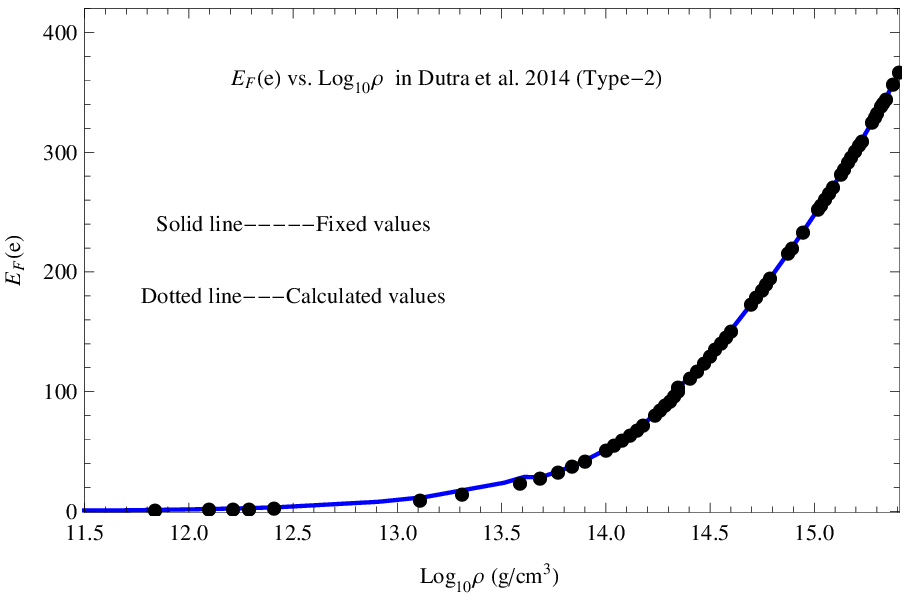}}\\
(a)&(b)\\
\end{tabular}
\end{center}
%\vspace*{8pt}
\caption{Numerically fitting Dutra et al. (2014)(Type-2).
 Left, the relation of $Y_{e}$ and $\rho$. Right, the relation of $E_{\rm F}(e)$
 and $\rho$. The range of $\rho$ is \textbf{(6.92$\times 10^{11}\sim 2.56\times 10^{15}$)}
 ~g~cm$^{-3}$.
 \label{fig10}}
\end{figure*}

Due to the restriction of space, the other six types in Dutra et al (2014) will no longer be considered.

\section{Comparisons and Discussions}\label{VI}
\subsection{Relations of $Y_{e}$ and $\rho$ in different models.}
The numerical simulations above indicate that the relation of $Y_{e}$ and $\rho$ is model dependent.
In order to see the differences between relations of $Y_{e}$ and $\rho$ in several simple
classical models (in this work) with those in RMF theory, we produce the diagrams of $Y_{e}$ vs.
$\rm Log_{10}\rho$ in the interior of a NS, as shown in Fig.11. In Fig.11 the black solid-line is obtained by
fitting in Sec.4, and the red solid-line is fitted from data of Tables 5-8 in Appendix B in this work. for Dutra et al. (2014)(Type-2). By comparing, some classical models (e.g., BPS and BBP models) may be superior to RMF theory model
in the low-density crustal region. For example, $Y_{e}$ in Dutra et al. (2014)(Type-2) begins to
appear at $\rho=6.92\times 10^{11}$~g~cm$^{-3}$, and always increases with $\rho$ in the high-density
range. However, employing the TMA parameter set, Dutra et al. (2014)(Type-2) will be more excellent than the simple
ideal $nep\mu$ and $nep\mu$ models when describing the relation of $Y_{e}$ and $\rho$ from EoSs, because
nucleon-nucleon interaction potentials are replaced by meson fields in RMF theory, rather than be ignored.
It is interesting that $Y_{e}$ in Dutra et al. (2014)(Type-2) grows with $\rho$ more quickly than in ideal $npe$ and $npe\mu$ systems.

In order to make a further comparison, in Fig. 11 we add one blue dot-line fitted
from Shen 2002\cite{Shen02}. In Shen (2002), the author
constructed the EoS in a wide NS density range using RMF theory. At low densities,
the Thomas-Fermi approximation was used to describe the nonuniform matter composed
of a lattice of heavy nuclei; while at high densities, the $TM1$ parameter set was adopted.
 Thus, $Y_{e}$ in  Shen (2002) firstly increases with $\rho$, then decreases with $\rho$.
 However, since the inclusion of hyperons softens the EoS considerably at high densities, the
maximum of the stellar mass in Shen (2002) is about 1.6\,$M_{Sun}$,(due to the depression of
hyperons on Fermions), which deviates from the observational NS mass limit of
 $\sim 1.97~M_{Sun}$ for PSR J1615-2230\cite{Demorest10}.

From the comparisons above, the differences of relations of $Y_{e}$ and $\rho$ between
different models are obvious. Despite these differences, we believe the simulations and comparisons presented
in this work will be useful in studying $E_{\rm F}(e)$ and $Y_{e}$ of a NS in the future.

\begin{figure*}[htb]
\centerline{\psfig{file=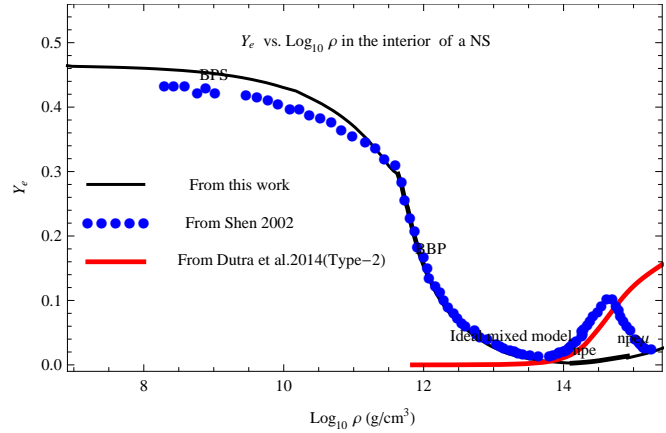,width=8.7cm}}
\vspace*{8pt}
\caption{$Y_{e}$ vs. $\rm Log_{10}\rho$ in the interior of a NS .
\protect \label{f11}}
\end{figure*}

\subsection{ Relation of $E_{\rm F}(e)$ and $n_{e}$}
Though both $E_{\rm F}(e)$ and $Y_{e}$ are definite functions of $\rho$.
the value of $E_{\rm F}(e)$ is ultimately determined by only parameter $n_{e}$,
\begin{eqnarray}
&&E_{\rm F}(e)= m_{e}c^{2}(1+ x_{e}^{2})^{1/2}\approx m_{e}c^{2}x_{e}
= m_{e}c^{2}(n_{e}3\pi^{2}\lambda_{e}^{3})^{1/3}\nonumber\\
 && =h c( \frac{3}{8\pi}n_{e})^{1/3}=\hbar c(3\pi^{2}n_{e})^{1/3}=6.12\times 10^{-11}n_{e}^{1/3}~({\rm MeV}), \label{38}
\end{eqnarray}
where $\hbar= h/2\pi$ is the reduced Plank¡¯s constant. Be note that this equation
does not depend on one certain matter model, i.e., Eq.(38) is also a general expression.
Based on Eq.(38), we plot a schematic diagram of $E_{\rm F}(e)$ vs. $n_{e}$ for
relativistic electrons in the whole interior of a common NS.
\begin{figure*}[htb]
\centerline{\psfig{file=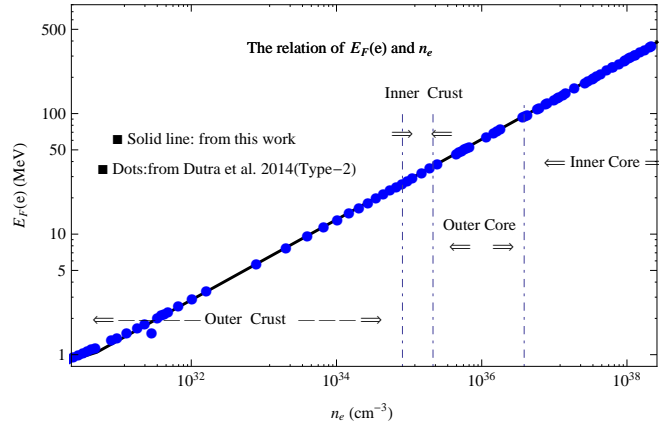,width=8.7cm}}
\vspace*{8pt}
\caption{The relation of $E_{\rm F}(e)$ and $n_{e}$ for relativistic electrons
in a common NS.
\protect \label{f12}}
\end{figure*}
% Since the electron Fermi energy $E_{\rm F}(e)$ increases with the electron number
% density $n_{e}$, and $n_{e}$ increases with  matter density $\rho$, $E_{\rm F}(e)$ also
% increases with matter density $\rho$ in the whole interior of a NS.

From Fig. 12, it is easy to see that, the larger the electron number density,
the bigger the electron Fermi energy become. In Fig.12, the dotted line is
fitted from data of Tables 5-7 in Dutra et al. (2014) (Type-2), while the solid line is obtained
from Eq.(38). These two lines are well in agreement with each other which indicates that
our calculations of Dutra et al. (2014)(Type-2) and the model self are surely correct.

Other Fermi parameters for electrons are also solely determined by $n_{e}$.
For example, the electron Fermi velocity, $v_{\rm F}(e)=\hbar k_{\rm F}/m_{e}=
\hbar(3\pi^{2}n_{e})^{1/3}/m_{e}$, the electron Fermi momentum, $p_{\rm F}(e)
=\hbar k_{\rm F}= \hbar(3\pi^{2}n_{e})^{1/3}$, and the Fermi kinetic energy of
relativistic electrons, $E_{K}^{\rm F}(e)\approx cp_{\rm F}(e)= c\hbar(3\pi^{2}
n_{e})^{1/3}$, due to $E_{K}^{\rm F}(e) \gg m_{e}c^2$), where $k_{\rm F}=
(3\pi^{2}n_{e})^{1/3}$ is the electron Fermi wave-vector. However, the relations
of $n_{e}$ and $\rho$ in different density regions of a NS are usually unknown, and the
known relations of $n_{e}$ and $\rho$ depend on EoS in some specific matter models.
Our study on $E_{\rm F}(e)$ may provide some conveniences in investigating EoS of a NS.

\subsection{Relations of $E_{\rm F}(e)$, $Y_{e}$ and $B$ }
In our previous studies, by introducing the Dirac $\delta$-function in superhigh magnetic fields
($B^{*}=B/B_{\rm cr}\gg 1$, $B_{\rm cr}=4.414\times 10^{13}$ G is the electron critical field),
we investigated the effects of strong magnetic fields on EoS and braking of magnetars\cite{Gao14,Gao16}.

Since superhigh magnetic fields can cause an increase in $Y_e$ by modifying the electron phase space, the
electron number density will increase with magnetic field strength $B$. From Table 1 of Gao et al.(2013)\cite{Gao13},
we obtain the diagrams of $Y_{e}$ vs. $\rho$ in BPS model in different magnetic fields, shown as in Fig. 13.
\begin{figure*}[htb]
\centerline{\psfig{file=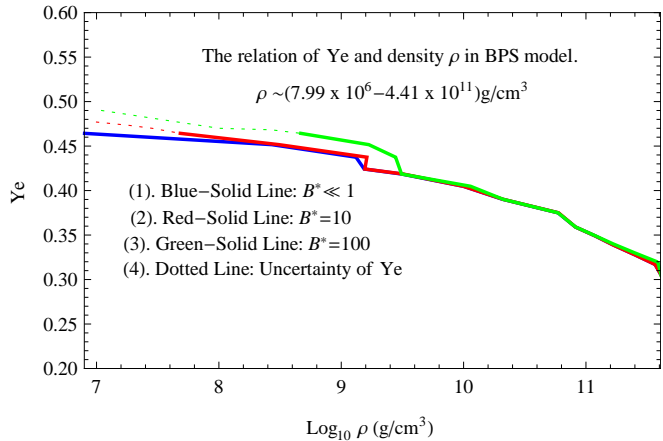,width=8.7cm}}
\vspace*{8pt}
\caption{The relation of $Y_{e}$ and $\rho$ in BPS model in different magnetic fields.
\protect \label{f13}}
\end{figure*}
Then, the Fermi energy of
electrons will increase with $B$. Very recently\cite{Gao15}, by introducing Landau level stability coefficient, we
obtain the relations of $E_{\rm F}(e)$, $Y_{e}$, $\rho$ (or $n_e$) and $B$,
\begin{eqnarray}
&&E_{\rm F}(e)=67.6\times\left(\frac{Y_e}{0.005647}\frac{\rho}{\rho_0}\right)^{1/3} \left(\frac{B}{B_{\rm cr}}\right)^{1/6}~~\rm MeV \nonumber\\
&&=6.86\times10^{-11}(n^{'}_{e})^{1/3}=6.86\times10^{-11}\left(\frac{B}{B_{\rm cr}}\right)^{1/6}(n_{e})^{1/3}~~\rm MeV, \label{39}
\end{eqnarray}
where $n^{'}_{e}$ is the number density of electrons in superhigh magnetic fields.
Numerically fitting the relations between $E_{\rm F}(e)$, $Y_{e}$ and
$\rho$ in superhigh magnetic fields will be considered in our future work.
\section{Conclusions}\label{VII}
In this paper we deduced a special solution to Fermi energy of relativistic
electrons in a common NS. By numerically simulating, we obtained several analytical formulae for
$Y_e$ vs. $\rho$ within classical matter models and Dutra et al.(2014)(Type-2).

As a representative model, BBP model is  selected to test
the validity of the specific solution to $E_{\rm F}(e)$. By comparing, our method of calculating
$E_{\rm F}(e)$ is more simple and convenient than that in BBP model, as well as those in other matter models.

Using the special solution to $E_{\rm F}(e)$, we can quickly and accurately calculate
the value of Fermi energy for relativistic electrons in any given matter density. The special
solution can be universally suitable for relativistic electrons regions in the circumstances of common NSs.

By numerically simulating, the special solution to $E_{\rm F}(e)$ has been proved
to be not only suitable for simple classical matter models, but also for and RMF
theory models), though Dutra et al.(2014)(Type-2) was only selected as a presented theory work.
Also, the special solution to $E_{\rm F}(e)$, as well as Eq.(38), could be
very useful in indirectly testing whether one EoS of a NS is correct, because Eq.(38) is the source
of the special solution of Eq.(14).

As an important parameter in EoS of a NS, the electron Fermi
energy is surely of very interest. The special solution to $E_{\rm F}(e)$
introduced by this work will be very useful in the future study on
EoS of NS¡¡matter under extreme conditions, though our methods of
treating EoS when numerically fitting are indeed simple.

\section*{Acknowledgments}
We sincerely thank the anonymous referee for carefully
reading the manuscript and providing valuable
comments that improved this paper substantially.
This work is supported by Xinjiang Natural Science Foundation No.2013211A053.

%\iffalse
\appendix
\section{Other classical matter models for the crust of a NS}
Since BPS model was developed on the basis of the semi-empirical
mass formula, it's necessary to introduce the work of Bethe, Borner
\& Sato (1970)\cite{Bethe70}(hereinafter¡°BBS model¡±)
that is a typical semi-empirical-mass-formular model. In BBS model, the energy per nucleon $E$ is given by
\begin{equation}
E=-c_{1}A+c_{2}Z^{2}A^{\frac{-1}{3}}+ c_{3}\frac{(N-Z)^2}{A}+c_{4}A^{\frac{2}{3}},
\label{A1}
 \end{equation}
where $N$ is the neutron number, $Z$ is the proton number, and $A=N+Z$ is
the nucleon number. The first term, the second term, the third term
and the fourth term on the right-hand side of Eq.(A1) denote
the volume energy, Coulomb energy, symmetry energy and  surface
energy, respectively; the energy constants $c_1$=16~MeV, $c_2$=0.72~MeV,
$c_3$=24~MeV, and $c_4$=18~MeV. The above equation fits the experimental
data very well for $A\sim 4$ to 260 after correcting for shell effects
and for Wigner term\cite{Bethe70}.
After ignoring the small neutron-proton mass difference, the electron chemical
potential $\mu_{\rm e}$ (i.e., the electron Fermi energy) becomes
 \begin{equation}
\mu_{\rm e}=\mu_{\rm n}-\mu_{\rm p}=\frac{\partial E}{\partial N}- \frac{\partial E}{\partial Z},
\label{A2}~
 \end{equation}
From Eq.(15) in BBS model\cite{Bethe70}, we get the relation of $A$ and $x$,
 \begin{equation}
A= \frac{c_{4}}{2c_{2}}x^{-2}=12.5 x^{-2}~.~
\label{A3}
\end{equation}
where $x=Z/A$. Thus, $Y_e$ is simply expressed as a function of $A$
\begin{equation}
~Y_{\rm e}\approx Z/A= (\frac{12.5}{A})^{1/2}~.
\label{A4}
\end{equation}
Based on Eq.(A4) and Table 2 in Section 4.1, we plot a schematic
diagram of $Y_e$ vs. $A$ for relativistic electrons in BBS
model and BPS model, as shown in Fig.9(a). Combining Eq.(A4) with
Table 1 in  BBS model\cite{Bethe70}, we plot a schematic diagram
of $E_{\rm F}(e)$ and $\rho$ in BBS model, as shown in Fig.9(b).
In this figure we add the fitting curve of $E_{\rm F}(e)$ vs. $Log_{10}(\rho)$ from BPS model\cite{BPS71}.
\begin{figure*}[htb]
\begin{center}
\begin{tabular}{cc}
\scalebox{0.65}{\includegraphics{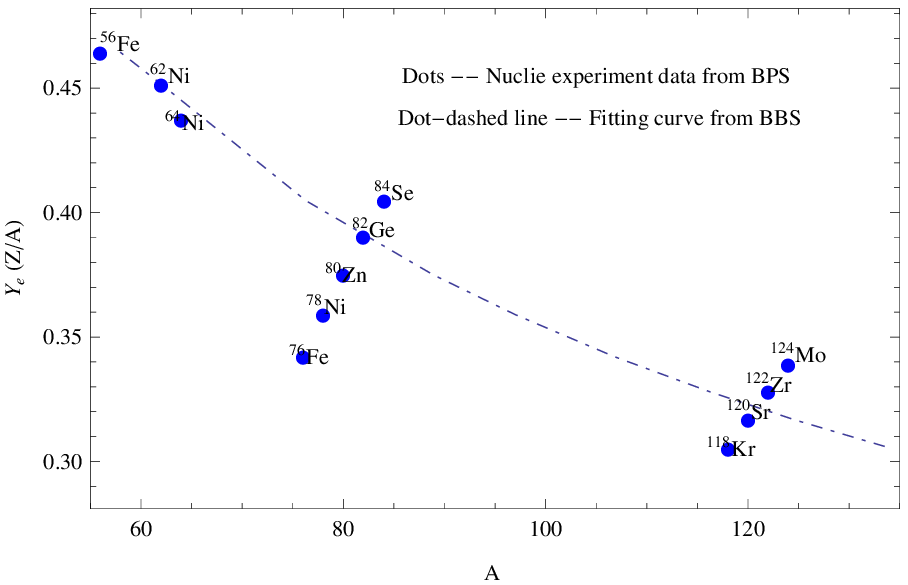}}&\scalebox{0.65}{\includegraphics{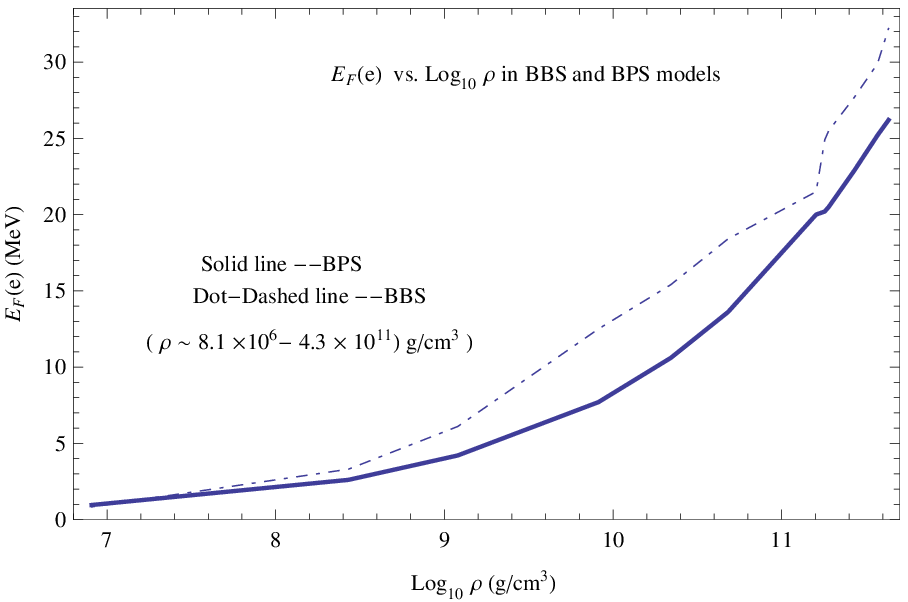}}\\
(a)&(b)\\
\end{tabular}
\end{center}
%\vspace*{8pt}
\caption{Comparisons of BBS model and BPS model in weak field limit. Left, the relations of $Y_e$ and $A$ in BBS and BPS models. Right, the relations of $E_{\rm F}(e)$ and $\rho$ in BBS and BPS models.
 \label{fig9}}
\end{figure*}

From Fig. 9, in BBS model the electron fraction $Y_e$ is solely
determined by nucleon number $A$, i.e., $Y_e$ monotonously
decreases with $A$; whereas in BPS model $Y_e$ is determined
by equilibrium nuclei and composition from EoS. The difference of $E_{\rm F}(e)$ between these two models increases with matter density.
Since the semi-empirical mass formula will not fit the experimental data when $\rho>\rho_d$,
we stop comparing these two models at the higher densities above $4.3\times 10^{13}$~g~cm$^{-3}$.
In addition, in BBS model the neutron drip density $\rho_d=2.8\times10^{11}$~g~cm$^{-3}$, and
the nuclei will disappear suddenly when $\rho=4.34\times10^{13}$~g~cm$^{-3}$. Due to the
introduction of lattice energy, BPS model is superior to
all the semi-empirical mass-formula models including BBS model.

Considering the limit of the semi-empirical mass formula, Buchler \& Barkat
(1971)\cite{Buchler71} (hereinafter ¡°BB model¡±) and Barkat, Buchler \& Wheeler (1972)\cite{Barkat72} (hereinafter ¡°BBW model¡±) calculated the nuclear composition and EOS, and gave the relation of $Z$ and $A$ and the expression of the surface energy by using Thomas-Fermi method.  However, both of these models cannot give definite values of $A$ and $A$ for stable nuclei, and the differences of surface-energy form between BB, BBW and BBP models are very large. Employing Hartree-Fock and Thomas-Fermi methods, Revenhall, Bennett \& Pethick (1972)\cite{Ravenhall72}(hereinafter ¡°RBP model¡±) modified the surface energy, calculated EoS, and gave the relation of $Z$ and $\rho$.  However, the expression of $Z$ and $\rho$ in RBW model is similar to that of BB model, but is different from that of BBP model.

\section{Partial calculations in Dura model(Type$-$2)}

\begin{table*}[htb]
\tbl{Partial calculations of  $n_{B}$, $n_{e}$, $Y_{e}$, $E_{\rm F}(e)$, $M$, $\varepsilon$ and $P$ in Dutra et al. 2014 (Type$-$2).}
{\begin{tabular}{@{}ccccccc@{}} \toprule
$n_{B}$ & $n_{e}$ &$Y_{e}$ &$E_{\rm F}(e)$ &$M$ & $\varepsilon$ & $P$ \\
$\rm fm^{-3}$ & $\rm cm^{-3}$  && MeV & $M_{\bigodot}$ &  Mev/fm$^{3}$ & Mev/fm$^{3}$\\
\colrule
$---$	&	$---$ &	$---$  &    $---$  &      $---$ & $---$ & $---$ \\
1.4$\times10^{-8}$   & 0  & 0	& 9.557$\times10^{-4}$	& 6.7628$\times10^{-5}$  & 7.167$\times10^{-8}$ & 3.301$\times10^{-14}$ \\
4.2$\times10^{-8}$   & 0  & 0	& 9.557$\times10^{-4}$ & 6.7628$\times10^{-5}$ & 7.167$\times10^{-8}$ & 3.301$\times10^{-14}$ \\
9.8$\times10^{-8}$	& 0	& 0	& 0.00425	& 0.0084 & 4.673$\times10^{-7}$  & 8.278$\times10^{-13}$ \\
1.12$\times10^{-7}$  & 0	& 0	& 0.00425	 & 0.0084 & 4.673$\times10^{-7}$   & 8.278$\times10^{-13}$ \\
%3.78$\times10^{-7}$ & 0	& 0	& 0.01026& 0.00378& 1.733$\times10^{-6}$	& 7.285$\times10^{-12}$ \\
5.88$\times10^{-7}$	 & 0& 0	& 0.01392 & 0.00301 & 2.731$\times10^{-6}$	& 1.547$\times10^{-11}$ \\
7.84$\times10^{-7}$ & 0	& 0	& 0.0157 & 0.00276 & 3.264$\times10^{-6}$	& 2.078$\times10^{-11}$ \\
%1.008$\times10^{-6}$ & 0& 0	& 0.01898& 0.00248& 4.331$\times10^{-6}$  & 3.316$\times10^{-11}$ \\
%3.808$\times10^{-6}$ & 0& 0	& 0.04921& 0.00279& 1.779$\times10^{-5}$	& 3.412$\times10^{-10}$ \\
6.72$\times10^{-6}$	  & 0	& 0	& 0.07279 & 0.0035& 3.171$\times10^{-5}$	& 8.820$\times10^{-10}$ \\
%8.176$\times10^{-6}$ & 0& 0	& 0.08333 & 0.00381 & 3.871$\times10^{-5}$	& 1.223$\times10^{-9}$ \\
9.856$\times10^{-6}$  & 0	& 0	& 0.09457& 0.00413 & 4.663$\times10^{-5}$	& 1.659$\times10^{-9}$ \\
1.312$\times10^{-5}$ & 0	& 0	& 0.11482  & 0.00467 & 6.202$\times10^{-5}$& 2.644$\times10^{-9}$ \\
3.78$\times10^{-4}$  & 5.2842$\times10^{30}$	& 1.398$\times10^{-5}$	& 1.1797	  & 0.02355  & 0.0018	& 5.736$\times10^{-7}$ \\
6.86$\times10^{-4}$  & 2.268$\times10^{31}$	& 3.306$\times10^{-5}$& 1.802& 0.02577 & 0.00327 & 1.4144$\times10^{-6}$ \\
8.96$\times10^{-4}$  & 4.198$\times10^{31}$	& 4.685$\times10^{-5}$& 2.1823 & 0.02700 & 0.00427 & 2.104$\times10^{-6}$ \\
0.00106	 & 6.202$\times10^{31}$	& 5.828$\times10^{-5}$& 2.4698 & 0.02783 & 0.0050  & 2.708$\times10^{-6}$ \\
0.00112& 6.964$\times10^{31}$& 6.218$\times10^{-5}$	& 2.563	& 0.02808  & 0.00533	& 2.918$\times10^{-6}$ \\
0.00123	  & 8.632$\times10^{31}$& 7.007$\times10^{-5}$	& 2.7459& 0.02855& 0.00587  & 3.350$\times10^{-6}$ \\
0.00134	 & 1.050$\times10^{32}$	& 7.810$\times10^{-5}$	& 2.924 & 0.02898  & 0.0064  & 3.797$\times10^{-6}$ \\
0.007& 4.331$\times10^{33}$	& 6.187$\times10^{-4}$	& 9.9637 & 0.06514  & 0.03339  & 3.121$\times10^{-5}$ \\
0.0112  & 1.286$\times10^{34}$	& 0.00115	& 14.310& 0.06183  & 0.05344  & 4.785$\times10^{-5}$ \\
0.0211	& 5.6976$\times10^{34}$	& 0.00271& 23.496 & 0.05927  & 0.10025  & 6.7902$\times10^{-5}$ \\
0.0322	& 1.6088$\times10^{35}$	& 0.005	   & 33.206	  & 0.05781  & 0.15377  & 9.6231$\times10^{-5}$ \\
0.0434 & 3.3639$\times10^{35}$  & 0.00775	 & 42.459 & 0.04904    & 0.2073   & 2.0973$\times10^{-4}$ \\
0.0546	& 5.9700$\times10^{35}$	& 0.01093& 51.405& 0.02952     & 0.26088  & 5.0427$\times10^{-4}$ \\
0.0602	& 7.6294$\times10^{35}$	& 0.01267& 55.784& 0.02714     & 0.2877   & 7.4956$\times10^{-4}$ \\
0.0714	  & 1.1725$\times10^{36}$	& 0.01642	        & 64.374	  & 0.04007     & 0.34142  & 0.00149 \\
0.0772	  & 1.4183$\times10^{36}$	& 0.01842	        & 68.589	  & 0.05169     & 0.36833  & 0.00201 \\
0.0882	  & 1.9963$\times10^{36}$	& 0.02263	        & 76.868	  & 0.08275     & 0.42229  & 0.00338 \\
0.0994	  & 2.6942$\times10^{36}$	& 0.0271	     & 84.946	  & 0.12331     & 0.47645  & 0.00526 \\
0.1051	  & 3.0893$\times10^{36}$	& 0.02942	    & 88.91	          & 0.14684     & 0.50362  & 0.00639 \\
0.1106	  & 3.5155$\times10^{36}$	& 0.03179	        & 92.825	  & 0.17234     & 0.53086  & 0.00767 \\
0.1218	  & 4.4622$\times10^{36}$	& 0.03664	      & 100.50	  & 0.22862     & 0.58554  & 0.01068 \\
0.1332	  & 5.5221$\times10^{36}$	& 0.04152	        & 107.90	  & 0.2904      & 0.64052  & 0.01427 \\
0.1386	  & 6.0665$\times10^{36}$	& 0.04377	        & 111.34	  & 0.32211     & 0.66813   & 0.01624 \\
0.1442	  & 6.6243$\times10^{36}$	& 0.04594	        & 114.65	  & 0.35433     & 0.69582   & 0.01835 \\
0.1498	  & 7.1972$\times10^{36}$	& 0.04805	        & 117.87	  & 0.38701     & 0.7236    & 0.0206 \\
0.1554	  & 7.7857$\times10^{36}$	& 0.0501	        & 120.99	  & 0.42008     & 0.75146   & 0.02299 \\
0.1611	  & 8.3895$\times10^{36}$	& 0.05211	        & 124.04	  & 0.45344     & 0.77941   & 0.02553 \\
0.1666	  & 9.0084$\times10^{36}$	& 0.05407	        & 127.02	  & 0.48699     & 0.80746   & 0.02822 \\
0.1722	  & 9.6420$\times10^{36}$	& 0.05599	        & 129.93	  & 0.52067     & 0.83559   & 0.03105 \\
0.1778	  & 1.0291$\times10^{37}$	& 0.05787	        & 132.78	  & 0.5544      & 0.86382   & 0.03404 \\
0.1834	  & 1.0951$\times10^{37}$	& 0.05971	        & 135.57	  & 0.58809     & 0.89215   & 0.03718 \\
0.1891	  & 1.1626$\times10^{37}$	& 0.06152	        & 138.30	  & 0.62169     & 0.92058   & 0.04047 \\
0.1946	  & 1.2314$\times10^{37}$	& 0.06328	        & 140.97	  & 0.65515     & 0.9491    & 0.04392 \\
0.2002	  & 1.3014$\times10^{37}$	& 0.06501	        & 143.59	  & 0.68839     & 0.9777    & 0.04752 \\
0.2058	  & 1.3725$\times10^{37}$	& 0.06669	        & 146.16	  & 0.72135     & 1.0065    & 0.05127 \\
0.2114	  & 1.4448$\times10^{37}$	& 0.06834	        & 148.68	  & 0.75401     & 1.0353    & 0.05518 \\
0.2171	  & 1.5181$\times10^{37}$	& 0.06996	        & 151.16	  & 0.78632     & 1.0642    & 0.05925 \\
0.2226	  & 1.5925$\times10^{37}$	& 0.07154	        & 153.59	  & 0.81824     & 1.0933    & 0.06347 \\
0.2282	  & 1.6678$\times10^{37}$	& 0.07309	        & 155.97	  & 0.84976     & 1.1224    & 0.06785 \\
0.2338	  & 1.7441$\times10^{37}$	& 0.07461	        & 158.31	  & 0.88083     & 1.1517    & 0.07238 \\
0.2492	  & 1.9583$\times10^{37}$	& 0.07858	        & 164.55	  & 0.96377     & 1.2328    & 0.08565 \\
0.2548	  & 2.0377$\times10^{37}$	& 0.07997	        & 166.74	  & 0.99298     & 1.2624    & 0.09077 \\
0.2716	  & 2.2802$\times10^{37}$	& 0.08395	        & 173.11	  & 1.07735     & 1.3522    & 0.10706 \\
0.2786	  & 2.3829$\times10^{37}$	& 0.08553	        & 175.67	  & 1.11105     & 1.3899    & 0.11426 \\
0.2884	  & 2.5283$\times10^{37}$	& 0.08767	        & 179.17	  & 1.15671     & 1.4431    & 0.12474 \\
0.2944	  & 2.6121$\times10^{37}$	& 0.08885	        & 181.13	  & 1.18203     & 1.4735    & 0.13094 \\
0.2996	  & 2.6965$\times10^{37}$	& 0.09011	        & 183.06	  & 1.20677     & 1.5041    & 0.13729 \\
0.3066	  & 2.8026$\times10^{37}$	& 0.09141	        & 185.43	  & 1.23689     & 1.5425    & 0.14544 \\
0.3122	  & 2.8880$\times10^{37}$	& 0.09251	        & 187.30	  & 1.26035     & 1.5734    & 0.15213 \\
0.3164	  & 2.9524$\times10^{37}$	& 0.09331	        & 188.68	  & 1.2776      & 1.5967    & 0.15725 \\
\botrule
\end{tabular} \label{ta5}}
\end{table*}

\begin{table*}[htb]
\tbl{Partial calculations of  $n_{B}$, $n_{e}$, $Y_{e}$, $E_{\rm F}(e)$, $M$, $\varepsilon$ and $P$ in Dutra et al.£¨2014£©(Type$-$2) (Continued).}
{\begin{tabular}{@{}ccccccc@{}} \toprule
$n_{B}$ & $n_{e}$ &$Y_{e}$ &$E_{\rm F}(e)$ &$M$ & $\varepsilon$ & $P$ \\
$\rm fm^{-3}$ & $\rm cm^{-3}$  && MeV & $M_{\bigodot}$ &  Mev/fm$^{3}$ & Mev/fm$^{3}$\\
\colrule
0.3206	  & 3.0169$\times10^{37}$	& 0.09410	        & 190.04	  & 1.29451     & 1.6201    & 0.16245 \\
0.3262	  & 3.1034$\times10^{37}$	& 0.09514	        & 191.84	  & 1.3166      & 1.6512    & 0.16952 \\
0.3318	  & 3.1902$\times10^{37}$	& 0.09615	        & 193.61	  & 1.33813     & 1.6825    & 0.17673 \\
0.336	  & 3.2555$\times10^{37}$	& 0.09689	        & 194.93	  & 1.3539      & 1.7061    & 0.18223 \\
0.3402	  & 3.3211$\times10^{37}$	& 0.09762	        & 196.23	  & 1.3694      & 1.7297    & 0.18782 \\
0.3444	  & 3.3868$\times10^{37}$	& 0.09834	        & 197.51	  & 1.38459     & 1.7534    & 0.19349 \\
0.3486	  & 3.4527$\times10^{37}$	& 0.09905	        & 198.78	  & 1.39947     & 1.7772    & 0.19924 \\
0.3556	  & 3.5631$\times10^{37}$	& 0.1002	        & 200.88	  & 1.42362     & 1.8171    & 0.20900 \\
0.3626	  & 3.6736$\times10^{37}$	& 0.10131	        & 202.94	  & 1.44698     & 1.857	    & 0.21898 \\
0.3696	  & 3.7847$\times10^{37}$	& 0.1024	        & 204.96	  & 1.46957     & 1.8971    & 0.22919 \\
0.3766	  & 3.8962$\times10^{37}$	& 0.10346	        & 206.95	  & 1.49138     & 1.9375    & 0.23961 \\
0.3836	  & 4.0081$\times10^{37}$	& 0.10449	        & 208.92	  & 1.51244     & 1.9781    & 0.25024 \\
0.3906	  & 4.1202$\times10^{37}$	& 0.10548	        & 210.85	  & 1.53275     & 2.0188    & 0.26108 \\
0.3976	  & 4.2328$\times10^{37}$	& 0.10646	        & 212.75	  & 1.55238     & 2.0598    & 0.27214 \\
0.4046	  & 4.3456$\times10^{37}$	& 0.10741	        & 214.62	  & 1.57132     & 2.1009    & 0.28341 \\
0.4186	  & 4.5721$\times10^{37}$	& 0.10922	        & 218.29	  & 1.60712     & 2.1838    & 0.30653 \\
0.4256	  & 4.6857$\times10^{37}$	& 0.11011	        & 220.08	  & 1.62408     & 2.2256    & 0.3184 \\
0.4326	  & 4.7996$\times10^{37}$	& 0.11095	        & 221.85	  & 1.6404      & 2.2675    & 0.33046 \\
0.4396	  & 4.9138$\times10^{37}$	& 0.11178	        & 223.61	  & 1.65612     & 2.3096    & 0.34271 \\
0.4466	  & 5.0281$\times10^{37}$	& 0.11259	        & 225.32	  & 1.67127     & 2.3521    & 0.35516 \\
0.4536	  & 5.1427$\times10^{37}$	& 0.11337	        & 227.02	  & 1.68583     & 2.3945    & 0.36779 \\
0.4606	  & 5.2574$\times10^{37}$	& 0.11414	        & 228.69	  & 1.69986     & 2.4372    & 0.38061 \\
0.4676	  & 5.3724$\times10^{37}$	& 0.11489	        & 230.35	  & 1.71337     & 2.4802    & 0.39361 \\
0.4746	  & 5.4875$\times10^{37}$	& 0.11562	        & 231.98	  & 1.72634     & 2.523     & 0.40679 \\
0.4886	  & 5.7183$\times10^{37}$	& 0.11703	        & 235.19	  & 1.7508      & 2.6101    & 0.43368 \\
0.4956	  & 5.8339$\times10^{37}$	& 0.11771	        & 236.76	  & 1.76233     & 2.6538    & 0.44738 \\
0.5026	  & 5.9497$\times10^{37}$	& 0.11838	        & 238.32	  & 1.7734      & 2.6977    & 0.46126 \\
0.5166	  & 6.1818$\times10^{37}$	& 0.11966           & 241.38	  & 1.79427     & 2.7861    & 0.4895 \\
0.5236	  & 6.2981$\times10^{37}$	& 0.12028	        & 242.88	  & 1.80409     & 2.8306    & 0.50387 \\
0.5446	  & 6.6477$\times10^{37}$	& 0.12207	        & 247.31	  & 1.83121     & 2.9652    & 0.54791 \\
0.5586	  & 6.8815$\times10^{37}$	& 0.12319	        & 250.16	  & 1.84751     & 3.0559    & 0.57803 \\
0.5656	  & 6.9985$\times10^{37}$	& 0.12374	        & 251.57	  & 1.85514     & 3.1015    & 0.59332 \\
0.5796	  & 7.2331$\times10^{37}$	& 0.12479	        & 254.35	  & 1.86949     & 3.1933    & 0.62432 \\
0.5852	  & 7.327$\times10^{37}$	& 0.12521	        & 255.45	  & 1.87489     & 3.2303    & 0.63688 \\
0.5964	  & 7.5152$\times10^{37}$	& 0.12601	        & 257.62	  & 1.88514     & 3.3045    & 0.66226 \\
0.6076	  & 7.7037$\times10^{37}$	& 0.12679	        & 259.75	  & 1.8947      & 3.3793    & 0.68800 \\
0.6216	  & 7.9397$\times10^{37}$	& 0.12773	        & 262.38	  & 1.90571     & 3.4734    & 0.72064 \\
0.6286	  & 8.0579$\times10^{37}$	& 0.12819	        & 263.67	  & 1.91086     & 3.5207    & 0.73716 \\
0.6356	  & 8.1762$\times10^{37}$	& 0.12864	        & 264.96	  & 1.91578     & 3.5682    & 0.7538 \\
0.6636	  & 8.6507$\times10^{37}$	& 0.13036	        & 269.99	  & 1.93329     & 3.7610    & 0.8216 \\
0.6776	  & 8.8887$\times10^{37}$	& 0.13118	        & 272.44	  & 1.94087     & 3.8571    & 0.85621 \\
0.6846	  & 9.0078$\times10^{37}$	& 0.13158	        & 273.65	  & 1.94439     & 3.9058    & 0.87369 \\
0.6916	  & 9.1271$\times10^{37}$	& 0.13197	        & 274.86	  & 1.94772     & 3.9548    & 0.89129 \\
0.6972	  & 9.2226$\times10^{37}$	& 0.13228	        & 275.81	  & 1.95028     & 3.9941    & 0.90544 \\
0.7028	  & 9.3182$\times10^{37}$	& 0.13259	        & 276.76	  & 1.95272     & 4.0335    & 0.91967 \\
0.7084	  & 9.4139$\times10^{37}$	& 0.13289	        & 277.71	  & 1.95507     & 4.073     & 0.93397 \\
0.7196	  & 9.6054$\times10^{37}$	& 0.13348	        & 279.58	  & 1.95947     & 4.1524    & 0.96277 \\
0.7266	  & 9.7253$\times10^{37}$	& 0.13385	        & 280.73	  & 1.96203     & 4.2023    & 0.9809 \\
0.7406	  & 9.9654$\times10^{37}$	& 0.13456	        & 283.03	  & 1.96675     & 4.3025    & 1.0175 \\
0.7476	  & 1.0086$\times10^{38}$	& 0.13491	        & 284.16	  & 1.96891     & 4.3529    & 1.0359 \\
0.7616	  & 1.0326$\times10^{38}$	& 0.13559	        & 286.40	  & 1.97286     & 4.4541    & 1.0731 \\
0.7686	  & 1.0447$\times10^{38}$	& 0.13592	        & 287.51	  & 1.97467     & 4.505	    & 1.0919 \\
0.7756	  & 1.0568$\times10^{38}$	& 0.13625	        & 288.62	  & 1.97634     & 4.5561    & 1.1107 \\
0.7826	  & 1.0689$\times10^{38}$	& 0.13658	        & 289.71	  & 1.97793     & 4.6073    & 1.1297 \\
0.7896	  & 1.0811$\times10^{38}$	& 0.1369	        & 290.80	  & 1.97941     & 4.6587    & 1.1487 \\
0.7966	  & 1.0931$\times10^{38}$	& 0.13722	        & 291.88	  & 1.98076     & 4.7103    & 1.1678 \\
\botrule
\end{tabular} \label{ta6}}
\end{table*}

\begin{table*}[htb]
\tbl{Partial calculations of  $n_{B}$, $n_{e}$, $Y_{e}$, $E_{\rm F}(e)$, $M$, $\varepsilon$ and $P$ in Dutra et al. (2014)(Type$-$2) (Continued).}
{\begin{tabular}{@{}ccccccc@{}} \toprule
$n_{B}$ & $n_{e}$ &$Y_{e}$ &$E_{\rm F}(e)$ &$M$ & $\varepsilon$ & $P$ \\
$\rm fm^{-3}$ & $\rm cm^{-3}$  && MeV & $M_{\bigodot}$ &  Mev/fm$^{3}$ & Mev/fm$^{3}$\\
\colrule
0.8148	  & 1.1246$\times10^{38}$	& 0.13802	        & 294.66	  & 1.98386     & 4.8451    & 1.2180 \\
0.8204	  & 1.1343$\times10^{38}$	& 0.13826	        & 295.51	  & 1.98471     & 4.8868    & 1.2335 \\
0.8246	  & 1.1416$\times10^{38}$	& 0.13844	        & 296.14       & 1.98530     & 4.9182    & 1.2452 \\
0.8316	  & 1.1538$\times10^{38}$	& 0.13874	        & 297.19	  & 1.98622     & 4.9706    & 1.2648 \\
0.8386	  & 1.1660$\times10^{38}$	& 0.13904	        & 298.23	  & 1.98704     & 5.0232    & 1.2845 \\
0.8456	  & 1.1781$\times10^{38}$	& 0.13933	        & 299.27	  & 1.98780     & 5.0759    & 1.3042 \\
0.8526	  & 1.1904$\times10^{38}$	& 0.13961	        & 300.30	  & 1.98847     & 5.1288    & 1.3240 \\
0.8596	  & 1.2026$\times10^{38}$	& 0.1399	        & 301.32	  & 1.98909     & 5.1819    & 1.3440 \\
0.8708	  & 1.2221$\times10^{38}$	& 0.14035	        & 302.95	  & 1.98991     & 5.2671    & 1.3762 \\
0.8764	  & 1.2319$\times10^{38}$	& 0.14057	        & 303.75	  & 1.99025     & 5.3099    & 1.3921 \\
0.8862	  & 1.2491$\times10^{38}$	& 0.14095	        & 305.16	  & 1.99075     & 5.3850    & 1.4204 \\
0.8918	  & 1.2589$\times10^{38}$	& 0.14117	        & 305.96	  & 1.99099     & 5.4280    & 1.4366 \\
0.8974	  & 1.2687$\times10^{38}$	& 0.14138	        & 306.75	  & 1.99117     & 5.4712    & 1.4529 \\
0.9156	  & 1.3007$\times10^{38}$	& 0.14206	        & 309.31	  & 1.99156     & 5.6122    & 1.5062 \\
0.9212	  & 1.3106$\times10^{38}$	& 0.14227	        & 310.09	  & 1.9916      & 5.6557    & 1.5227 \\
0.9268	  & 1.3204$\times10^{38}$	& 0.14247 	        & 310.86	  & 1.99162     & 5.6994    & 1.5393 \\
0.9380	  & 1.3402$\times10^{38}$	& 0.14288	        & 312.40	  & 1.99153     & 5.7871    & 1.5725 \\
0.9436	  & 1.3501$\times10^{38}$	& 0.14308	        & 313.17	  & 1.99144     & 5.8311    & 1.5892 \\
0.9492	  & 1.3600$\times10^{38}$	& 0.14328	        & 313.93	  & 1.99132     & 5.8752    & 1.6060 \\
0.9548	  & 1.3699$\times10^{38}$	& 0.14347	        & 314.69	  & 1.99119     & 5.9194    & 1.6228 \\
0.9604	  & 1.3798$\times10^{38}$	& 0.14367	        & 315.45	  & 1.99101     & 5.9637    & 1.6396 \\
0.9660	  & 1.3897$\times10^{38}$	& 0.14386	        & 316.20	  & 1.99082     & 6.0080    & 1.6565 \\
0.9716	  & 1.3996$\times10^{38}$	& 0.14405	        & 316.96	  & 1.99059     & 6.0525    & 1.6734 \\
0.9772	  & 1.4096$\times10^{38}$	& 0.14424	        & 317.70	  & 1.99033     & 6.0971    & 1.6904 \\
0.9856	  & 1.4245$\times10^{38}$	& 0.14453	        & 318.82	  & 1.98991     & 6.1641    & 1.7159 \\
0.9912	  & 1.4344$\times10^{38}$	& 0.14472	        & 319.56	  & 1.98958     & 6.2090    & 1.7330 \\
0.9968	  & 1.4444$\times10^{38}$	& 0.14490	        & 320.30	  & 1.98925     & 6.2539    & 1.7502 \\
1.0024	  & 1.4543$\times10^{38}$	& 0.14509	        & 321.03	  & 1.98889     & 6.2989    & 1.7673 \\
1.0080	  & 1.4643$\times10^{38}$	& 0.14527	        & 321.77	  & 1.98851     & 6.3440    & 1.7846 \\
1.0360	  & 1.5143$\times10^{38}$	& 0.14617	        & 325.38	  & 1.98628     & 6.5710    & 1.8713 \\
1.0416	  & 1.5243$\times10^{38}$	& 0.14634	        & 326.10	  & 1.98578     & 6.6167    & 1.8888 \\
1.0472	  & 1.5343$\times10^{38}$	& 0.14652	        & 326.81	  & 1.98525     & 6.6625    & 1.9063 \\
1.0528	  & 1.5444$\times10^{38}$	& 0.14669	        & 327.52	  & 1.98470     & 6.7083    & 1.9239 \\
1.0640	  & 1.5645$\times10^{38}$	& 0.14703	        & 328.94	  & 1.98359     & 6.8003    & 1.9591 \\
1.0696	  & 1.5745$\times10^{38}$	& 0.14721	        & 329.64	  & 1.98299     & 6.8465    & 1.9768 \\
1.0752	  & 1.5846$\times10^{38}$	& 0.14737	        & 330.34	  & 1.98239     & 6.8927    & 1.9945 \\
1.0808	  & 1.5946$\times10^{38}$	& 0.14754	        & 331.04	  & 1.98177     & 6.9391    & 2.0123 \\
1.0864	  & 1.6047$\times10^{38}$	& 0.14771	        & 331.74	  & 1.98114     & 6.9855    & 2.0301 \\
1.0920	  & 1.6148$\times10^{38}$	& 0.14787	        & 332.43	  & 1.98049     & 7.0320    & 2.0479 \\
1.0976	  & 1.6249$\times10^{38}$	& 0.14804	        & 333.12	  & 1.97981     & 7.0786    & 2.0658 \\
1.1032	  & 1.6350$\times10^{38}$	& 0.14820	        & 333.81	  & 1.97915     & 7.1253    & 2.0837 \\
1.1088	  & 1.6451$\times10^{38}$	& 0.14837	        & 334.50	  & 1.97845     & 7.1721    & 2.1017 \\
1.1228	  & 1.6704$\times10^{38}$	& 0.14877	        & 336.20	  & 1.97668     & 7.2895    & 2.1467 \\
1.1270	  & 1.6780$\times10^{38}$	& 0.14889	        & 336.71	  & 1.97613     & 7.3248    & 2.1603 \\
1.1396	  & 1.7008$\times10^{38}$	& 0.14924	        & 338.23	  & 1.97444     & 7.4311    & 2.2011 \\
1.1438	  & 1.7084$\times10^{38}$	& 0.14936	        & 338.73	  & 1.97386     & 7.4666    & 2.2148 \\
1.1480	  & 1.7160$\times10^{38}$	& 0.14948	        & 339.23	  & 1.97328     & 7.5022    & 2.2285 \\
1.1648	  & 1.7465$\times10^{38}$	& 0.14994	        & 341.23	  & 1.97091     & 7.6450    & 2.2834 \\
1.1704	  & 1.7567$\times10^{38}$	& 0.15009	        & 341.89	  & 1.97009     & 7.6928    & 2.3017 \\
1.1760	  & 1.7669$\times10^{38}$	& 0.15024	        & 342.55	  & 1.96926     & 7.7406    & 2.3202 \\
1.1816	  & 1.7771$\times10^{38}$	& 0.15040	        & 343.21	  & 1.96843     & 7.7886    & 2.3386 \\
1.1872	  & 1.7873$\times10^{38}$	& 0.15055	        & 343.87	  & 1.96759     & 7.8366    & 2.3571 \\
1.1928	  & 1.7975$\times10^{38}$	& 0.15069	        & 344.52	  & 1.96675     & 7.8847    & 2.3756 \\
1.1984	  & 1.8077$\times10^{38}$	& 0.15084	        & 345.17	  & 1.96589     & 7.9330    & 2.3942 \\
\botrule
\end{tabular} \label{ta7}}
\end{table*}

\begin{table*}[htb]
\tbl{\textbf{Partial calculations of  $n_{B}$, $n_{e}$, $Y_{e}$, $E_{\rm F}(e)$, $M$, $\varepsilon$ and $P$ in Dutra et al. (2014)(Type$-$2)} (Continued).}
{\begin{tabular}{@{}ccccccc@{}} \toprule
$n_{B}$ & $n_{e}$ &$Y_{e}$ &$E_{\rm F}(e)$ &$M$ & $\varepsilon$ & $P$ \\
$\rm fm^{-3}$ & $\rm cm^{-3}$  && MeV & $M_{\bigodot}$ &  Mev/fm$^{3}$ & Mev/fm$^{3}$\\
\colrule
1.2026	  & 1.8154$\times10^{38}$	& 0.15095	        & 345.66	  & 1.96525     & 7.9692    & 2.4081 \\
1.2068	  & 1.8230$\times10^{38}$	& 0.15106	        & 346.15	  & 1.96460     & 8.0054    & 2.4220 \\
1.2110	  & 1.8307$\times10^{38}$	& 0.15117	        & 346.63	  & 1.96394     & 8.0418    & 2.4360 \\
1.2152	  & 1.8384$\times10^{38}$	& 0.15128	        & 347.11	  & 1.96327     & 8.0781    & 2.4501 \\
1.2194	  & 1.8461$\times10^{38}$	& 0.15139	        & 347.60	  & 1.96261     & 8.1145    & 2.4640 \\
1.2236	  & 1.8538$\times10^{38}$	& 0.15150	        & 348.08	  & 1.96195     & 8.1510    & 2.4781 \\
1.2278	  & 1.8614$\times10^{38}$	& 0.15161	        & 348.56	  & 1.96127     & 8.1875    & 2.4921 \\
1.2320	  & 1.8691$\times10^{38}$	& 0.15172	        & 349.04	  & 1.96061     & 8.2241    & 2.5062 \\
1.2362	  & 1.8768$\times10^{38}$	& 0.15182	        & 349.52	  & 1.95992     & 8.2607    & 2.5203 \\
1.2404	  & 1.8845$\times10^{38}$	& 0.15193	        & 349.99	  & 1.95925     & 8.2973    & 2.5344 \\
1.2446	  & 1.8922$\times10^{38}$	& 0.15204	        & 350.47	  & 1.95855     & 8.3340    & 2.5486 \\
1.2488	  & 1.8999$\times10^{38}$	& 0.15214	        & 350.95	  & 1.95786     & 8.3708    & 2.5627 \\
1.2530	  & 1.9077$\times10^{38}$	& 0.15225	        & 351.42	  & 1.95717     & 8.4076    & 2.5769 \\
1.2572	  & 1.9154$\times10^{38}$	& 0.15235	        & 351.89	  & 1.95648     & 8.4444    & 2.5911 \\
1.2614	  & 1.9231$\times10^{38}$	& 0.15246	        & 352.37	  & 1.95578     & 8.4813    & 2.6053 \\
1.2656	  & 1.9308$\times10^{38}$	& 0.15256	        & 352.84	  & 1.95507     & 8.5182    & 2.6195 \\
1.2698	  & 1.9385$\times10^{38}$	& 0.15266	        & 353.31	  & 1.95437     & 8.5552    & 2.6338 \\
1.2740	  & 1.9463$\times10^{38}$	& 0.15277	        & 353.78	  & 1.95366     & 8.5922    & 2.648 \\
1.2782	  & 1.9540$\times10^{38}$	& 0.15287	        & 354.24	  & 1.95296     & 8.6293    & 2.6623 \\
1.2824	  & 1.9617$\times10^{38}$	& 0.15297	        & 354.71	  & 1.95224     & 8.6665    & 2.6766 \\
1.2866	  & 1.9695$\times10^{38}$	& 0.15308	        & 355.18	  & 1.95153     & 8.7036    & 2.6909 \\
1.2908	  & 1.9772$\times10^{38}$	& 0.15318	        & 355.64	  & 1.95079     & 8.7408    & 2.7053 \\
1.2950	  & 1.9850$\times10^{38}$	& 0.15328	        & 356.11	  & 1.95008     & 8.7781    & 2.7196 \\
1.2992	  & 1.9927$\times10^{38}$	& 0.15338	        & 356.57	  & 1.94935     & 8.8154    & 2.7340 \\
1.2782	  & 1.9540$\times10^{38}$	& 0.15287	        & 354.24	  & 1.95296     & 8.6293    & 2.6623 \\
1.2992	  & 1.9927$\times10^{38}$	& 0.15338	        & 356.57	  & 1.94935     & 8.8154    & 2.7340 \\
1.3034	  & 2.0005$\times10^{38}$	& 0.15348	        & 357.03	  & 1.94862     & 8.8528    & 2.7484 \\
1.3076	  & 2.0082$\times10^{38}$	& 0.15358	        & 357.49	  & 1.94789     & 8.8902    & 2.7628 \\
1.3118	  & 2.0160$\times10^{38}$	& 0.15368	        & 357.95	  & 1.94718     & 8.9276    & 2.7772 \\
1.3160	  & 2.0238$\times10^{38}$	& 0.15378	        & 358.41	  & 1.94644     & 8.9651    & 2.7917 \\
1.3202	  & 2.0315$\times10^{38}$	& 0.15388	        & 358.87	  & 1.94570     & 9.0027    & 2.8062 \\
1.3244	  & 2.0393$\times10^{38}$	& 0.15398	        & 359.33	  & 1.94497     & 9.0403    & 2.8206 \\
1.3286	  & 2.0471$\times10^{38}$	& 0.15408	        & 359.78	  & 1.94423     & 9.0779    & 2.8351 \\
1.3328	  & 2.0549$\times10^{38}$	& 0.15418	        & 360.24	  & 1.94347     & 9.1156    & 2.8497 \\
1.3370	  & 2.0627$\times10^{38}$	& 0.15427	        & 360.69	  & 1.94274     & 9.1533    & 2.8642 \\
1.3412	  & 2.0704$\times10^{38}$	& 0.15437	        & 361.14	  & 1.94200     & 9.1911    & 2.8788 \\
1.3454	  & 2.0782$\times10^{38}$	& 0.15447	        & 361.60	  & 1.94125     & 9.2289    & 2.8933 \\
1.3496	  & 2.0860$\times10^{38}$	& 0.15457	        & 362.05	  & 1.94051     & 9.2668    & 2.9079 \\
1.3538	  & 2.0938$\times10^{38}$	& 0.15466	        & 362.50	  & 1.93976     & 9.3047    & 2.9225 \\
1.3580	  & 2.1016$\times10^{38}$	& 0.15476	        & 362.95	  & 1.93897     & 9.3427    & 2.9372 \\
1.3622	  & 2.1094$\times10^{38}$	& 0.15485	        & 363.40	  & 1.93823     & 9.3807    & 2.9518 \\
1.3664	  & 2.1172$\times10^{38}$	& 0.15495	        & 363.85	  & 1.93749     & 9.4187    & 2.9665 \\
1.3706	  & 2.1251$\times10^{38}$	& 0.15505	        & 364.29	  & 1.93673     & 9.4568    & 2.9811 \\
1.3748	  & 2.1329$\times10^{38}$	& 0.15514	        & 364.74	  & 1.93598     & 9.4949    & 2.9958 \\
1.3790	  & 2.1407$\times10^{38}$	& 0.15523	        & 365.18	  & 1.93523     & 9.5331    & 3.0105 \\
1.3832	  & 2.1485$\times10^{38}$	& 0.15533	        & 365.63	  & 1.93446     & 9.5713    & 3.0253 \\
1.3888	  & 2.1589$\times10^{38}$	& 0.15545	        & 366.22	  & 1.93344     & 9.6224    & 3.0449 \\
1.3944    & 2.1694$\times10^{38}$	& 0.15558	        & 366.81	  & 1.93242     & 9.6735    & 3.0646 \\
1.40	  & 2.1798$\times10^{38}$	& 0.15570	        & 367.40	  & 1.9314      & 9.7247    & 3.0844 \\
\botrule
\end{tabular} \label{ta8}}
\end{table*}
%\fi

\begin{thebibliography}{99}
\bibitem{Xu02} R.X. Xu, {\it Astrophys. J. Lett.} {\bf 570} (2002) 65.
\bibitem{Du09} Y.J. Du, et al., {\it Mon. Not. R. Astron. Soc.} {\bf 399}(2009) 1587.
\bibitem{Lai13}X.Y. Lai, C.Y. Gao and R.X. Xu, 2013,{\it Mon. Not. R. Astron. Soc.} {\bf 431} (2013) 3290.
\bibitem{Yakovlev01} D.G. Yakovlev, A.D. Kaminker, O.Y. Gnedin and et al. {\it Phys. Rep.} {\bf 354}(2001) 1.
\bibitem{Gao11a} Z.F. Gao, N. Wang, J.P. Yuan and et al., {\it Astrophys. Space Sci.} {\bf 333}(2011a) 427.
\bibitem{Gao11b} Z.F. Gao, Q.H. Peng, N. Wang and et al., {\it Astrophys. Space Sci.} {\bf 336} (2011b) 427.
\bibitem{Gao12a} Z.F. Gao, Q.H. Peng, N. Wang and et al., {\it Astrophys. Space Sci.} {\bf 342}(2012a) 55.
\bibitem{Wang12} K. Wang, Z.Q. Luo and  Y.L. Li, {\it Chin. Phys. Lett.} {\bf 29}(2012) 049701 .
\bibitem{Liu13}J.J. Liu, {\it Astrophys. Space Sci.} {\bf 347} (2013), 117.
\bibitem{Shapiro83} S. L. Shapiro and S. A. Teukolsky, {\it Black Holes, White Drarfs, and Neutron Stars}, (New York, Wiley-Interscience, 1983).
\bibitem{Lattimer04}J.M. Lattimer and M.Prakash, {\it Sceience}, {\bf 304}(2004) 536.
\bibitem{Gao12b}Z.F. Gao, Q.H. Peng, N. Wang and et al., {\it Chin. Phys. B} {\bf 21}(2012b) 057109.
\bibitem{Gao13} Z.F. Gao, N. Wang, Q.H. Peng and et al.,{\it Mod. Phys. Lett A.} {\bf 28(36)} (2013) 1350138.
\bibitem{Gao11c} Z.F. Gao, N. Wang, D.L. Song and et al., {\it Astrophys. Space Sci.} {\bf 334} (2011c) 281.
\bibitem{Dutra14} M. Dutra, O. Lourenco, S.S. Avancini and et al.,{\it Phys. Rev. C.} {\bf 90} (2014), 055203.
\bibitem{Horowitz04} C. J. Horowitz, et al.,{\it Phys. Rev. C.} {\bf 69} (2004), 045804.
\bibitem{Owen05} B. J. Owen, {\it Phys. Rev. Lett.} {\bf 95} (2005) 21101.
\bibitem{Steiner05} A. W. Steiner, M. Prakash, J.M. Lattimer and et al.,{\it Phys. Rep.} {\bf 410} (2005) 325.
\bibitem{Burrows06} A. Burrows, S. Reddy and T. A. Thompson, {\it Nucl. Phys. A.} {\bf 777} (2006) 356.
\bibitem{BPS71} G. Baym, C. Pethick and P. Sutherland, {\it Astrophys. J.} {\bf 170}(1971) 299.
\bibitem{Ruster06}S.B. Ruster, M. Hempel and J. Schaffnet-Bielich, {\it Phys.Rev. C.} {\bf 73}, 035804 (2006).
\bibitem{Tsuruta09}S. Tsuruta, J. Sadino, A. akatsuka and  et al., {\it Astrophys. J.} {\bf 691} (2009) 621.
\bibitem{BBP71} G. Baym, H.A. Bethe and C.J. Pethick, {\it Nuclear Phys. A.} {\bf 175} (1971) 225.
\bibitem{Canuto74}V. Canuto, {\it Ann. Rev. Astron. Astrophys.} {\bf 12} (1974) 167.
\bibitem{Salpeter61} E.E. Salpeter, {\it Astrophys. J.} {\bf 134} (1961) 669.
\bibitem{Lattimer01} J. M. Lattimer and M. Prakash, {\it Astrophys. J.} {\bf 550}, 426 (2001).
\bibitem{Lattimer07} J. M. Lattimer and M. Prakash, {\it Phys. Rep.}, {\bf 442}, 109 (2007).
\bibitem{NV73}L.W. Negele and D. Vautherin, {\it Nucl. Phys. A.} {\bf 207} (1973), 298 .
\bibitem{Atta14} D. Atta and D.N. Basu,{\it Phys. Rev. C.} {\bf 90}, 035802 (2014).
\bibitem{Lattimer91}J.M. Lattimer, M. Prakash, C.J. Pethick and et al., {\it Phys. Rev. Lett.}
        {\bf 66} (1991), L2701.
\bibitem{Haensel07} P. Haensel, A.Y. Potekhin and D.G. Yakovlev, {\it Neutron Stars I, Equation of
State and Structure}, (New York, Springer, 2007).
\bibitem{Xu13a}Y. Xu, G.Z. Liu, H.Y. Wang and et al., {\it Chin. Phys. Lett.} {\bf 29}(2013a) 059701.
\bibitem{Xu13b}Y. Xu, G.Z. Liu, C.Z. Liu and et al., {\it Chin. Phys. Lett.} {\bf 30} (2013b) 129501.
\bibitem{Glendenning85}N. K. Glendenning, {\it Astrophys. J.} {\bf 293} (1985) 470
\bibitem{Schaffner93} J. Schaffner, C. B. Dover, A. Gal and et al., {\it Phys Lett B.}{\bf 71} (1993) 1328.
\bibitem{Shen02} H. Shen {\it Phys. Rev. C.} {\bf 65} (2002), 035802.
\bibitem{Yang08} F. Yang, H. Shen, {\it Phys. Rev. C.} {\bf 77} (2008), 025801.
\bibitem{Dutra12} M. Dutra {\it Phys. Rev. C.} {\bf 85} (2012), 035201.
\bibitem{Demorest10} P. B. Demorest, et al. {\it Nature (London)} {\bf 467} (2010) 1081.
\bibitem{Sugahara94}Y. Sugahara and H. Toki, {\it Nucl. Phys. A.} {\bf 579} (1994), 557.
\bibitem{Geng05}L. Geng, et al. {\it Progress of Theoretic Physics} {\bf 113} (2005), 785.
\bibitem{Singh12} D. Singh, et al.{\it Int. J. Mod. Phys. E.} {\bf 21} (2012), 1250076.
%\bibitem{Lalazissis97}G. A. Lalazissis, et al. {\it Phys. Rev. C.} {\bf 55} (1997) 540.
\bibitem{Lattimer13}J. Lattimer and Y. Lim, {\it Astrophys. J. Lett.} {\bf 771} (2013), 51.
\bibitem{Gao14} Z.F. Gao, X.J. Zhao, D.L. Song and et al., {\it Astron. Nachr.} {\bf 335, No.6/7} (2014) 653.
\bibitem{Gao16}Z.F. Gao, X.-D. Li, N. Wang, et al. {\it Mon. Not. R. Astron. Soc.} {\bf 456} (2016),55
\bibitem{Gao15}Z.F. Gao,N. Wang, Y. Xu, et al., {\it Astron. Nachr.}{\bf 336, No.8/9} (2015) 866.
%\bibitem{Xu05}R.X. Xu, {\it Mon. Not. R. Astron. Soc.} {\bf 356}(2005), 359.
%\bibitem{Becker09} W. Becker,(ed.), {\it Neutron stars and pulsars}. (Astrophys. Space Sci. % Library, 2009) p.~357.
%\bibitem{Pandharipande71} V.R. Pandharipande, {\it Nucl. Phys. A.} {\bf 178} (1971) 123.
%\bibitem{Freedman74} D.Z. Freedman, {\it Phys. Rev. D.} {\bf 9} (1974) 1389.
\bibitem{Bethe70} H.A. Bethe, G. Borner and K. Sato, {\it Astron.\& Astrophys.} {\bf 7}(1970) 279 .
\bibitem{Buchler71} J.R. Buchler and Z. Barkat, {\it Astrophys. J. Lett.} {\bf 7}(1971) L179 .
\bibitem{Barkat72} Z. Barkat, J. R. Buchler and  J. C. Wheeler, {\it Astrophys. J.} {\bf 173} (1972) 183.
\bibitem{Ravenhall72}D.G. Ravenhall, C.D. Bennett and C.J. Pethick, {\it Phys. Rev. Lett.} {\bf 28} (1972) L978.
\end{thebibliography}
\end{document}